\documentclass[12pt]{article}
\pdfoutput=1

\usepackage{jheppub}
\usepackage{graphicx,amsmath,amsfonts,amssymb,hyperref,cleveref,siunitx,color,braket}
\usepackage{cancel,enumerate,appendix,multirow}

\definecolor{niceblue}{rgb}{0.388235, 0.627451, 0.847059}
\definecolor{nicered}{rgb}{0.7,0.1,0.1}
\definecolor{nicegreen}{rgb}{0.1,0.5,0.1}

\hypersetup{colorlinks, citecolor=nicegreen, linkcolor=nicered}
\usepackage{array}
\newcolumntype{C}[1]{>{\centering\arraybackslash}p{#1}}

\newcommand\pfrac[2]{\frac{\partial #1}{\partial #2}}

\newcommand{\dderiv}[2]{{\mathrm d#1}/{\mathrm d#2}}
\newcommand\DB{\operatorname{DB}}

\newcommand\Nc{N_\text{cells}}
\newcommand\NE{N_\text{Ebins}}

\renewcommand\d{\mathrm d}
\renewcommand{\arraystretch}{1.1}

\allowdisplaybreaks

\preprint{INT-PUB-21-031, N3AS-21-015, IFT-UAM/CSIC-21-134}

\title{Statistical significance of the sterile-neutrino hypothesis in the context of reactor and gallium data}

\author[a,b,c]{Jeffrey M. Berryman,}
\affiliation[a]{Department of Physics, University of California, Berkeley, CA 94720, USA}
\affiliation[b]{Department of Physics and Astronomy, University of Kentucky, Lexington, KY 40506, USA}
\affiliation[c]{Institute for Nuclear Theory, University of Washington, Seattle, WA 98195, USA}
\emailAdd{jeffberryman@berkeley.edu}

\author[d]{Pilar Coloma,}
\affiliation[d]{Instituto de F\'{i}sica Te\'{o}rica UAM-CSIC, Calle Nicol\'{a}s Cabrera 13-15, 28049 Madrid, Spain}
\emailAdd{pilar.coloma@ift.csic.es}

\author[e]{Patrick Huber,}
\affiliation[e]{Center for Neutrino Physics, Physics Department, Virginia Tech, Blacksburg, VA 24061, USA}
\emailAdd{pahuber@vt.edu}

\author[f]{Thomas Schwetz,}
\affiliation[f]{Institut f\"{u}r Astroteilchenphysik, Karlsruher Institut f\"{u}r Technologie
(KIT), D-76021 Karlsruhe, Germany}
\emailAdd{schwetz@kit.edu}

\author[f]{Albert Zhou}
\emailAdd{albert.zhou@kit.edu}

\abstract{
We evaluate the statistical significance of the 3+1 sterile-neutrino hypothesis using
$\nu_e$ and $\bar\nu_e$ disappearance data from reactor, solar and gallium radioactive
source experiments. Concerning the latter, we investigate the implications of the
recent BEST results. For reactor data we focus on relative measurements independent of
flux predictions. For the problem at hand, the usual $\chi^2$-approximation to
hypothesis testing based on Wilks' theorem has been shown in the literature to be
inaccurate. We therefore present results based on Monte Carlo simulations, and find
that this typically reduces the significance by roughly $1\,\sigma$ with respect to the
na\"ive expectation. We find no significant indication in favor of
sterile-neutrino oscillations from reactor data. On the other hand, gallium data
(dominated by the BEST result) show more than $5\,\sigma$ of evidence supporting the
sterile-neutrino hypothesis, favoring oscillation parameters in agreement with
constraints from reactor data. This explanation is, however, in significant
tension ($\sim 3\,\sigma$) with solar neutrino experiments. In order to assess the
robustness of the signal for gallium experiments we present a discussion of the impact
of cross-section uncertainties on the results.
}

\begin{document}

\maketitle

\newpage
\section{Introduction}

There is a number of long-standing experimental results in neutrino physics that defy Standard
Model (SM) expectations, and at least individually find an explanation by new oscillations
with a mass-squared splitting of around 1\,eV$^2$; for reviews see {\em e.g.,}
refs.~\cite{Boser:2019rta,Dasgupta:2021ies}. This scale is much larger than the mass-squared
splittings associated with the three neutrinos of the SM, and hence points to the existence of
a forth neutrino, which must be a singlet under the SM gauge group ({\em i.e.,\ }sterile), in
order to agree with the invisible decay width of the $Z$-boson~\cite{ALEPH:2005ab}. A sterile
neutrino would open a fermionic portal to a dark sector and, as such, is well motivated.

In 2011, a new piece of evidence appeared in the form of the so-called \emph{reactor
antineutrino anomaly}~\cite{Mention:2011rk}: new evaluations of the prediction of the
antineutrino flux from a reactor~\cite{Mueller:2011nm,Huber:2011wv} resulted in a 6\% increase
of the expected neutrino signal. This implied that past experiments (which were in agreement
with previous calculations) were in fact observing a deficit of the same magnitude. In
parallel, renewed analyses of radiochemical source-calibration measurements done for the
solar-neutrino experiments SAGE~\cite{Abdurashitov:1998ne,Abdurashitov:2005tb} and
GALLEX~\cite{Hampel:1997fc, Kaether:2010ag} also indicated a deficit compatible with the
result from reactors: the so-called \emph{gallium anomaly}~\cite{Laveder:2007zz,Acero:2007su,Giunti:2010zu}.
Thus, for the first time, there was considerable evidence of anomalous $\nu_e$ and $\bar\nu_e$
disappearance.

Because of this, the past decade has seen a plethora of experimental efforts (both with
reactor neutrinos and radioactive sources), as well as theoretical efforts in nuclear physics,
to improve source and cross-section predictions. On the experimental side, a number of reactor
experiments at short distances (6 -- 25\,m) from the reactor core have been conducted:
DANSS~\cite{DANSS-2021}, NEOS~\cite{NEOS:2016wee}, PROSPECT~\cite{Andriamirado:2020erz},
STEREO~\cite{AlmazanMolina:2019qul} and Neutrino-4~\cite{Serebrov:2020kmd}. These experiments
derive their results for sterile-neutrino oscillations by comparing the measured neutrino
spectra at different distances from the reactor, and thus their results do not rely on
reactor-flux modeling. While some of these reported evidence favoring the sterile-neutrino
hypothesis, others did not, and thus the question of the statistical compatibility of the
different data sets and their interpretation in the form of a global fit is
relevant~\cite{Dentler:2017tkw,Gariazzo:2018mwd}. All of these experiments look for an
oscillatory distortion in the ratio of measured spectra, and there are a number of subtleties
that arise in their statistical
interpretation~\cite{Agostini:2019jup,Giunti:2020uhv,PROSPECT:2020raz,Coloma:2020ajw}, which
render the usual approach based on Wilks' theorem~\cite{Wilks} unreliable.

In parallel, in order to test the gallium anomaly, the BEST collaboration recently presented a
new result of a two-zone gallium radiochemical measurement employing a chromium-51
source~\cite{Barinov:2021asz}. Although they did not observe a different rate in the two
volumes, their results confirmed the earlier deficit with much smaller error bars. This new
result brings the gallium anomaly to a new level, making it highly significant and raising the
question of its compatibility with other data \cite{Barinov:2021mjj,Giunti:2021kab}.

Since 2011, significant effort has gone into improving our understanding of reactor
antineutrino fluxes, both from a theoretical perspective and from many measurements of
beta-feeding functions of individual isotopes. These efforts have led to a much better
understanding of the underlying physics, see {\em e.g.,\ }ref.~\cite{Hayes:2016qnu}, but have
not resolved the reactor antineutrino anomaly. Global analyses of past reactor neutrino
data~\cite{Giunti:2017yid, Giunti:2019qlt,Berryman:2020agd}, precise neutrino-spectrum
measurements~\cite{DayaBay:2021dqj} and fuel-evolution measurements~\cite{DayaBay:2017jkb}
have been indicating for a while that there might be a problem with the prediction for
uranium-235. The recent absolute flux determination from a reactor with only uranium-235
fission by the STEREO experiment~\cite{STEREO:2020fvd} confirms this. The basis for the
Huber-Mueller reactor fluxes are integral beta-spectrum measurements conducted in the 1980s at
the Institute Laue-Langevin
(ILL)~\cite{VonFeilitzsch:1982jw,Schreckenbach:1985ep,Hahn:1989zr}. Recently, the ratio of
uranium-235 to plutonium-239 integral beta spectra was measured~\cite{Kopeikin:2021ugh},
indicating the possibility that the normalization of the original ILL data for uranium-235 was
too high. Taking this new information into account, a global flux
analysis~\cite{Giunti:2021kab} finds that there is no evidence for the reactor antineutrino
anomaly in total rates. All data sets and methods of prediction agree in total rate if one
replaces the normalization of uranium-235 with the new result of ref.~\cite{Kopeikin:2021ugh}.
Therefore, any support for the sterile-neutrino hypothesis from reactor neutrino data now
would have to come from the observation of an actual oscillation by DANSS, NEOS, PROSPECT,
STEREO or Neutrino-4. Note that the spectrum discrepancy between flux models and reactor
neutrino measurements~\cite{DayaBay:2015lja,DoubleChooz:2015mfm,RENO:2015ksa} around 5\,MeV of
neutrino energy, colloquially known as the 5-MeV bump, has no bearing on the question of a
sterile neutrino for these experiments, since in taking the ratio of spectra measured at
different distances, the bump effectively cancels.

Finally, a number of electron-neutrino disappearance results from accelerator experiments may
also be interpreted in the context of a sterile neutrino. Specifically, both KARMEN and
LSND~\cite{Armbruster:1998uk,LSND:2001fbw,Conrad:2011ce} have provided bounds from a
charged-current reaction on $^{12}$C. A measurement of the T2K experiment reported in
ref.~\cite{T2Kdisapp}, its main purpose being a determination of the charged-current
electron-neutrino cross section and characterization of the $\nu_e$ beam component, also leads
to a mild indication of disappearance. A somewhat similar result appeared recently based on
MicroBooNE data~\cite{Denton:2021czb}, see also ref.~\cite{Arguelles:2021meu}. All of these
results are not very statistically significant, as of yet, and do not strongly favor
oscillation over no-oscillation nor vice versa; we therefore do not include them in our fits.
Similar remarks apply to the tritium beta-decay experiment KATRIN \cite{KATRIN:2020dpx}, whose
exclusion depends strongly on whether the light neutrino masses can be neglected with respect
to the mass of the much heavier fourth neutrino \cite{KATRIN:2020dpx,Giunti:2019fcj}.

Let us also remark that additional eV-scale neutrinos face severe constraints from cosmology,
both in regard to their early-time contributions as relativistic degrees of freedom and their
late-time contributions to the energy density as hot dark matter, suppressing the formation of
small-scale structure, see {\it e.g.,} ref.~\cite{Hagstotz:2020ukm} for a recent global
analysis and refs.~\cite{Boser:2019rta,Dasgupta:2021ies} for reviews and further references. However, any cosmological bound relies on the assumption of thermalization of the sterile neutrino in
the early universe and may thus be evaded in an alternative cosmology. Our goal here is to
present results that are robust with respect to cosmological assumptions.

In this paper, we present a global fit of up-to-date results from the DANSS, NEOS, PROSPECT,
STEREO and Neutrino-4 reactor experiments, the past gallium results from SAGE and GALLEX, as
well as the recent BEST result and solar-neutrino results in the framework of one additional
sterile neutrino (the so-called 3+1-oscillation hypothesis). The results presented are
independent of reactor-flux predictions and are thus complementary to the results of
ref.~\cite{Giunti:2021kab}. Since Wilks' theorem can not be used reliably
here~\cite{Agostini:2019jup, Giunti:2020uhv, PROSPECT:2020raz,Coloma:2020ajw}, we perform a
full Feldman-Cousins analysis~\cite{Feldman:1997qc}.

The analysis results are presented in the main text; individual experiments, as well as details of the
Feldman-Cousins analysis, are described in the appendices. \Cref{sec:reactor} contains a
description of the reactor data, as well as the results of their combination. In
\cref{sec:gallium}, we discuss the gallium experiments and the dependence of their results
with the charged-current cross section for neutrino capture. In \cref{sec:results-global}, we
evaluate the statistical compatibility of the various data sets, paying particular attention
to the compatibility between solar and gallium data, and present our global-fit results. In
\cref{sec:conclusions}, we present a summary of our findings and draw our conclusions.

\section{Reactor experiments}
\label{sec:reactor}%

\subsection{Description of the data used} \label{sec:exp-reactor}%

In the following, we provide a brief summary of the reactor experiments whose data we have
used; detailed descriptions of our analyses --- including the precise forms of the $\chi^2$
functions and our techniques for generating pseudo data --- are presented in
\cref{app:exp-impl}.%

\bigskip
\textbf{DANSS}
At DANSS, the detector is located on a movable platform under an industrial $3.1~\rm{GW}_{th}$
reactor~\cite{DANSS-2021}. It measures the neutrino spectrum at three baselines: bottom,
middle and top ($I \equiv B, M, T$, respectively). The DANSS collaboration presents their data
with the following ratios
\begin{align}
\label{eq:DANSS-R}
	R_1^i = \frac{n_B^i}{n_T^i} && R_2^i = \frac{n_M^i}{\sqrt{n_B^i n_T^i}},
\end{align}
where $i$ indicates the energy bin and \(n_I^i = N_I^i/\Delta t_I\). Here, \(N_I^i\) stands
for the event rates the energy bin $i$ and \(\Delta t_I\) is the corresponding exposure time
period for each detector location $I$. A total of 36~energy bins are used in our analysis,
spanning energies between \numrange{1.5}{6} \si{MeV}. Exposure times and the
background-subtracted rates for the top baseline were extracted from ref.~\cite[p.~12]
{DANSS-2021}. Doing this, we get a total event rate for the top baseline of $\sum_i n_T^i =
4132.61$ events per day, for energies between \numrange{1.5}{6} \si{MeV} (in prompt energy).
Details on our \(\chi^2\) implementation can be found in \cref{app:DANSS}.%

\bigskip
\textbf{NEOS}
The NEOS detector consists of approximately 1000~L of homogeneous 0.5\% Gd-loaded liquid
scintillator. It is located at about 23.7~m from a commercial reactor core with
a thermal power of 2.8 GW in the Hanbit nuclear-power complex in Yeonggwang, South Korea. The
detector operated for 46 days with the reactor off and 180 days with the reactor on. NEOS
measured $1976.7 \pm 3.4$ inverse-beta-decay candidates per day (with prompt energies between
1 and 10 MeV) during the reactor-on period, as opposed to $85.1 \pm 1.4$ events in the
reactor-off period. For our analysis, we take the data from
ref.~\cite[fig.\ 3a]{NEOS:2016wee}: the black points are assumed to be the
background-subtracted counts and are normalized to the quoted total signal rate,
\(1976.7-85.1=1891.6\) events/day.

The NEOS collaboration does a shape-only analysis using the neutrino spectrum from
ref.~\cite{DayaBay:2016ssb} plus corrections due to different fission fractions (see
\cref{app:NEOS} for details); therefore, we leave the normalization completely free in our fit. We build a binned
$\chi^2$, with 61 energy bins\footnote{Note that the last bin is
\emph{not} included in the total rate stated in the paper, but {\em is} used in the analysis.}
between \numrange1{10} \si{MeV}.

\bigskip
\textbf{PROSPECT}
The PROSPECT experiment uses a large, highly segmented detector. Segments are grouped into 10
different baselines (see fig.\ 39 of ref.~\cite{Andriamirado:2020erz}). Data are then reported
as 10 energy spectra --- one for each cluster of segments --- with 16 energy bins. Our
analysis benefits from the thorough data release accompanying ref.~\cite{Andriamirado:2020erz}, including the background-subtracted spectra for each baseline bin and the covariance matrix describing them, which includes both statistical and systematic uncertainties. The $\chi^2$ is calculated following ref.~\cite[eq.\ 11]{Andriamirado:2020erz} (see also \cref{app:PROSPECT}).%

\bigskip
\textbf{STEREO}
The STEREO detector contains six cells, each with a different effective baseline. The
experiment was divided up into two phases: for each cell, phase-I data are presented in ten
bins from 1.625 to 6.625 MeV of prompt energy, while phase-II data are presented in eleven
bins from 1.625 to 7.125 MeV (in both cases the bin sizes are set to 0.5 MeV). The data have
been published alongside ref.~\cite{AlmazanMolina:2019qul}; however, the raw event rates have
been normalized to the prediction under the no-oscillation hypothesis (after minimization over
systematic uncertainties). Therefore, we also rescaled our predictions in order to compare to
their data, using as the nuisance parameters (for the prediction under the no-oscillation
hypothesis) the values from fig.\ 31 of ref.~\cite{AlmazanMolina:2019qul}. The STEREO
analysis contains a total of 40 pulls: we are able to minimize 14 of these analytically, while
the remaining 26 are minimized numerically. Further details on our implementation of STEREO
can be found in \cref{app:STEREO}.

\bigskip
\textbf{Neutrino-4}
Our analysis of Neutrino-4~\cite{Serebrov:2020kmd} is based on the previous analysis performed
in ref.~\cite{Coloma:2020ajw}. Neutrino-4 also uses a segmented detector. The data are binned
using nine 0.5-MeV energy bins from 2.3 to \SI{6.8}{MeV}, and twenty-four 0.235-m baseline
bins from 6.25 to \SI{11.89}m. The resulting 216 bins in \(L/E\) are then clustered into
consecutive groups of 8 bins, leading to a total of 27 data points. The analysis is performed
using the ratios
\begin{align} \label{eq:R-nu4}
	R_i = \frac{d_i}{\frac{1}{N}\sum_j d_j},
\end{align}
where $d_i$ is the number of events in the $i^\text{th}$ bin. In our analysis, we use the
first 19 bins for the combined phase-1 and phase-2 data sets (the blue points in fig.\ 47 of
the v2 arXiv preprint of ref.~\cite{Serebrov:2020kmd}). Note that although the event counts
$d_i$ are uncorrelated, the ratios $R_i$ are correlated through the denominator in
\cref{eq:R-nu4}. However, we have estimated these correlations and find that they do not change our results, so they will be neglected here. Further details on our $\chi^2$
implementation and analysis can be found in \cref{app:nu4} as well as in
ref.~\cite{Coloma:2020ajw}. %

\subsection{Results of the reactor analysis}
\label{sec:results-reactors} %

\begin{table}[t!]
	\centering
	\setlength{\tabcolsep}{3.5pt}
	\begin{tabular}{|l|ccc|cccc|}
		\hline
		&\(\chi^2_\text{min}\)/dof&\(\Delta m^2_\text{min}\)&\(\sin^22\theta_\text{min}\) &
		\(\Delta\chi^2_{3\nu}\)& $p_0$ & $\#\sigma$ & $\#\sigma^{(W)}$ \\
		\hline
		\textbf{DANSS}	&			& 1.3 \si{eV^2}	& 0.014	& 3.2 &			&	& 1.3 \\
		Our Fit			& 77.6/70	& 1.32 \si{eV^2}& 0.011	& 3.6 &	43.8\%	&0.8& 1.4 \\
		\hline
		\textbf{NEOS}	& 57.5/59	& 1.73 \si{eV^2}& 0.05	& 6.5 &	22\%	&1.2& 2.1 \\
		Our Fit			& 59/58		& 2.95 \si{eV^2}& 0.16	& 7.4 &	19.4\%	&1.3& 2.2 \\
		\hline
		\textbf{PROSPECT}&119.3/142	& 1.78 \si{eV^2}& 0.11	& 4.0 &	57\%	&0.6& 1.5 \\
		Our Fit			& 118.1/158	& 1.75 \si{eV^2}& 0.11	& 4.3 &	63.3\%	&0.5& 1.6 \\
		\hline
		\textbf{STEREO}	& 128.4/112	& 8.95 \si{eV^2}& 0.63	& 9.0 &	 9\%	&1.7& 2.5 \\
		Our Fit			& 128.6/112	& 8.72 \si{eV^2}& 0.59	& 7.8 &	15.8\%	&1.4& 2.3 \\
		\hline
		\textbf{Nu-4}	& 14.7/17	& 7.26 \si{eV^2}& 0.38	& 15.3&			&	& 3.5\\
		Our Fit			& 16.1/17	& 7.31 \si{eV^2}& 0.38	& 12.6&	1.5\%	&2.4& 3.1\\
		\hline
		\textbf{REACTORS}	& 428/421 & \SI{8.86}{eV^2} & 0.26 & 7.3 &27.4\%& 1.1	& 2.2 \\
		\textbf{W/ Solar}	& 432/425 & \SI{1.30}{eV^2} & 0.014& 6.6 &17.8\%& 1.3	& 2.1 \\
		\textbf{W/ Gallium}\footnote{The reactor-with-gallium \(p_0\) values and significances are
		extrapolations and not MC calculations. The lower $p$-value assumes a maximum-Gauss
		distribution with \(n=60\) \cite{Coloma:2020ajw}, the upper $p$-value corresponds to a
		linear extrapolation of the tail of the distribution; see \cref{fig:null-hyp}.}
		& 433/427 & \SI{8.86}{eV^2} & 0.32 & 38.8 &
		\multirow{2}{1.85cm}{$(0.14\to1.4)\times10^{-7}$} & \(5.7\to5.3\) & 5.9 \\[3ex]
		\hline
	\end{tabular}
	\caption{Comparison of our fit results with those from the individual collaborations
	(where available) for DANSS \cite[p.\ 17]{DANSS-2021}, NEOS \cite{NEOS:2016wee}, PROSPECT
	\cite[§VIII.C]{Andriamirado:2020erz}, STEREO \cite[table 5]{AlmazanMolina:2019qul} and
	Neutrino-4 \cite[§21 of v2 of arXiv preprint]{Serebrov:2020kmd}. The first three columns provide the
	$\chi^2$ value at the global minimum and the corresponding oscillation parameters. The
	remaining columns show the $\Delta\chi^2$ of the null hypothesis (no new
	oscillations), the $p$-value of the null hypothesis calculated from MC simulations ($p_0$),
	and its equivalent number of Gaussian standard deviations, \(\#\sigma\). The
	last column shows the corresponding number of standard deviations when Wilks' theorem is
	assumed, {\em i.e.,\ }assuming $\Delta\chi^2_{3\nu}$ is distributed as $\chi^2(2\text
	{ dof})$. We also show our combined results for all reactors, as well as reactors with
	solar, and reactors with gallium data. The gallium analysis uses the Kostensalo
	{\em et al.\ }cross section \cite{Kostensalo:2019vmv}.
	}
	\label{tab:best-fit}
\end{table}%

\begin{figure}[t]
	\centering
	\includegraphics[width=\textwidth]{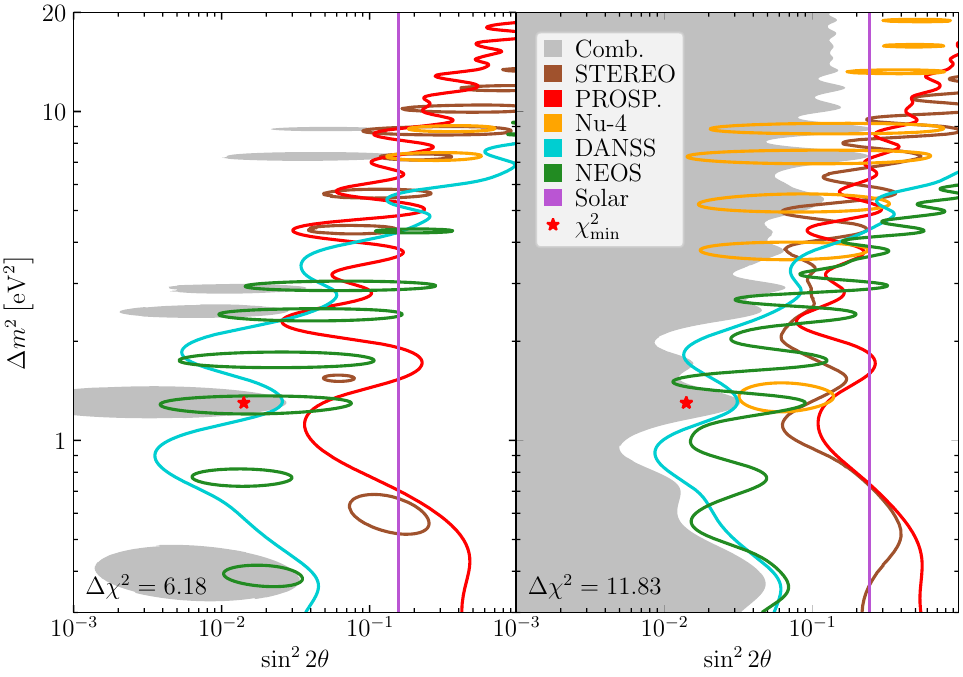}
	\caption{Isocontours of \(\Delta\chi^2=6.18\) (left) and 11.83 (right) for reactor
	experiments and the solar constraint, corresponding to 2, \(3\sigma\) levels under Wilks'
	theorem, respectively. The grey regions correspond to the combined analysis and the
	red star is the best-fit point of all data combined.
	}
	\label{fig:chisq-all}
\end{figure}%

For all the reactor data listed above, we can reproduce the results of the corresponding
collaborations with good accuracy when adopting the same assumptions and statistical method.
This is shown in \cref{tab:best-fit}, which compares our best-fit points to the ones from the
respective collaborations, showing reasonable agreement in all cases. The \(\chi^2\) values of
the global minimum ($\chi^2_\text{min}$) are also shown: as can be seen, our results match
very well those obtained by the different collaborations in all cases. Notice the
slightly different number of dof between our analysis and the collaboration's for the NEOS and
PROSPECT analyses, which is due to a different number of nuisance parameters used in the fit.

In the right-most part of the table, we evaluate the data's preference for the presence
of sterile-neutrino oscillations by considering the test statistic\footnote{This test
statistic has been denoted by $T$ in ref.~\cite{Coloma:2020ajw}.}
\begin{equation}\label{eq:T}
	\Delta \chi^2_{3\nu} = \chi^2_{3\nu} - \chi^2_\text{min},
\end{equation}
where $\chi^2_{3\nu}$ is the $\chi^2$ value for the null hypothesis, $\sin^2 2\theta = 0$
($\theta$ being the angle that parametrizes the mixing between the sterile neutrino and the
light states), and $\chi^2_\text{min}$ is the $\chi^2$ value at the best-fit point. We use
this test statistic to evaluate the $p$-value of the null hypothesis in the following way.
For a given experiment, the distribution that $\Delta \chi^2_{3\nu}$ follows can be obtained
from Monte Carlo (MC) simulations: a set of pseudo experiments is randomly generated according
to the expected statistical fluctuations of event rates in each bin (see \cref{app:MC} for
details). The obtained distribution of the test statistic can then be compared to the
experimental value in order to obtain the corresponding $p$-value. The $p$-values we obtain,
as well as the corresponding number of Gaussian standard deviations, are given in
\cref{tab:best-fit}. Again, the comparison with the collaborations' corresponding results
shows good agreement.

If Wilks' theorem held, then $\Delta \chi^2_{3\nu}$ would follow a $\chi^2$ distribution with
2~dof (corresponding to the two parameters, $\sin^22\theta$ and $\Delta m^2$, which we
minimize over). In the last column of \cref{tab:best-fit}, we show the number of standard
deviations corresponding to that assumption, $\#\sigma^{(W)}$. Comparing with the previous
column, we find that Wilks' theorem consistently overestimates the significance by about one
standard deviation, confirming previous studies \cite{Agostini:2019jup,Giunti:2020uhv,
PROSPECT:2020raz,Coloma:2020ajw}. Therefore, in order to assess the significance for the
presence of sterile neutrinos from these data, it is essential to study the distribution of
the test statistic numerically.%

\begin{figure}[t]
	\centering
	\includegraphics[width=0.9\textwidth]{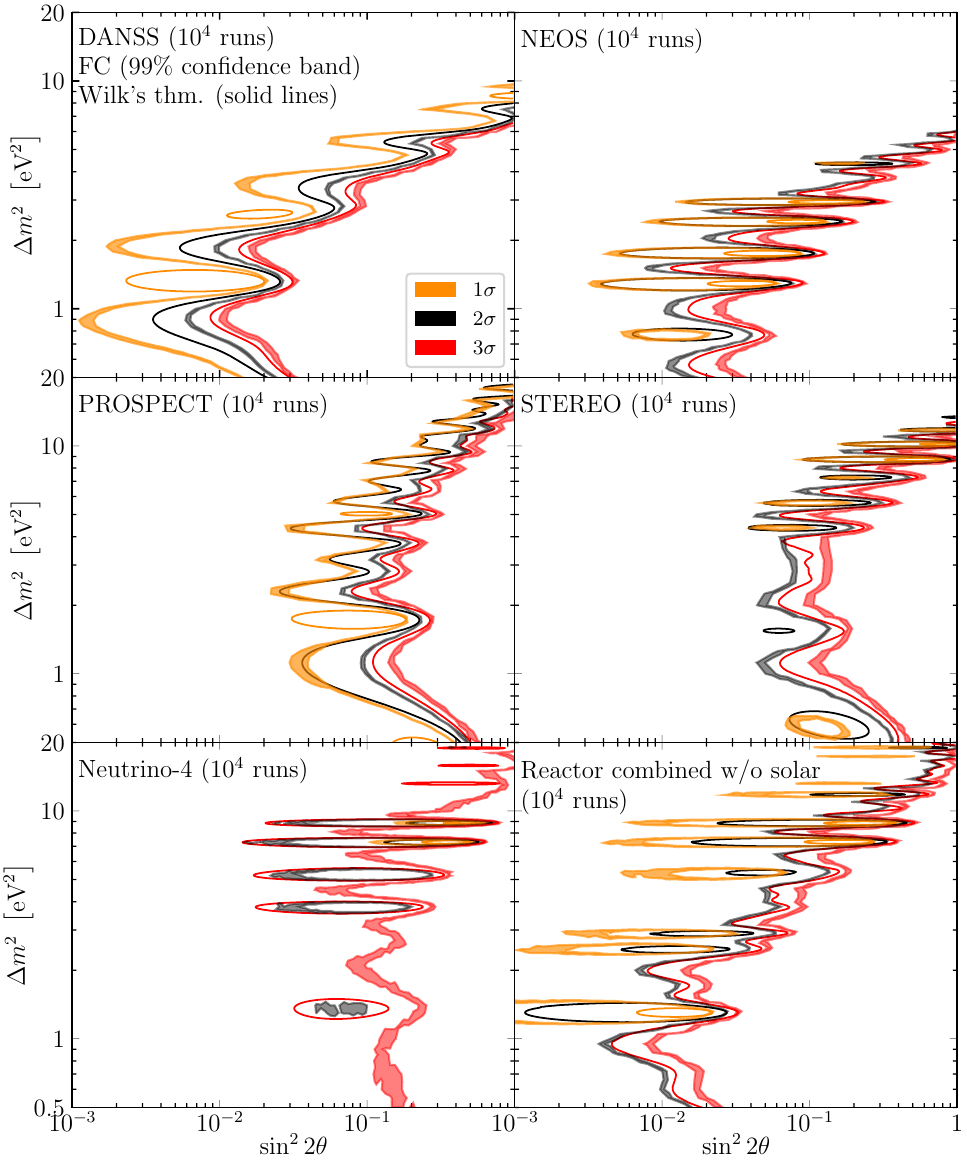}
	\caption{Confidence regions at 68.27\%, 95.45\%, 99.73\%~CL (1, 2, 3$\sigma$, respectively)
	for the individual reactor experiments and all reactors combined (bottom right). The bands
	correspond to the Feldman-Cousins analysis, where their width indicates the 99\% confidence
	spread of the limit due to the finite size of our MC sample (see \cref{app:FC-error}). Thin
	curves are obtained under the assumption of Wilks' theorem (assuming that $\Delta\chi^2$
	follows a $\chi^2$ distribution with 2~dof).
	}
	\label{fig:FCMC-expmnts}
\end{figure} %

Next, let us discuss the differences in the allowed confidence regions obtained in the
$\sin^22\theta$-$\Delta m^2$ plane. \Cref{fig:chisq-all} shows the contours for constant
$\Delta \chi^2 = 6.18$ and 11.83 for the various reactor experiments individually (as well as
their combination) together with the constraint from solar neutrinos (see below), where
\begin{equation}\label{eq:Dchisq}
	\Delta\chi^2(\sin^2 2\theta, \Delta m^2) =
	\chi^2(\sin^2 2\theta, \Delta m^2) - \chi^2_\text{min} \,.
\end{equation}
If Wilks' theorem were valid, then these contours would correspond to 95.45\% and 99.73\%
confidence regions, respectively. However, since Wilks' theorem cannot be applied here, the
confidence regions should be determined from simulation, according to the Feldman-Cousins (FC)
prescription~\cite{Feldman:1997qc} as follows. For a given pair of assumed true values of
$\sin^2 2\theta$ and $\Delta m^2$, MC simulations of statistical fluctuations in the data can
be used to determine the expected distribution of $\Delta \chi^2$ as defined in
\cref{eq:Dchisq} and compared to the value measured experimentally, in order to determine the
confidence level (CL) at which a given point of parameter space can be rejected. The full
allowed region in the $\sin^22\theta$-$\Delta m^2$ plane is then obtained by repeating this
procedure for all points in parameter space. Confidence regions obtained in this way have the
correct coverage by construction.

The colored bands in \cref{fig:FCMC-expmnts} show such FC confidence regions, corresponding to
1, 2, and 3$\sigma$ significance, for the individual reactor experiments (as well as for their
combination, in the last panel), where the width of the bands indicates the 99\% spread due to
the finite size of the MC sample (see \cref{app:FC-error} for details). Comparing these bands
to the lines of the same color, which correspond to $\Delta\chi^2$ contours assuming that
Wilks' theorem applies, we find that in most cases the true CL is reduced approximately by one
Gaussian standard deviation, similar to the null-hypothesis case. Also, we do not find closed
confidence regions at 99.73\%~CL for any of the reactor experiments, in agreement with the
$p$-values of the null hypothesis reported in \cref{tab:best-fit}.%

Although none of the individual experiments shows a clear signal at high significance, several
of them favor the sterile-neutrino-oscillation hypothesis at a mild confidence level,
$\lesssim 2\sigma$. Therefore, the question arises, whether these indications add up and lead
to a significant result in a combined analysis. We saw already in \cref{fig:chisq-all} that
the allowed regions for the different experiments do not show a consistent overlapping pattern
and therefore an enhancement of the various hints is not expected from their combination. This
is confirmed and quantified by the results of our global reactor MC analysis, reported in
\cref{tab:best-fit} as well as in the bottom-right panel of \cref{fig:FCMC-expmnts}: the
combined reactor fit gives $\Delta\chi^2_{3\nu} = 7.3$ with a $p$-value corresponding to
$1.1\sigma$. Hence, we find that reactor data are statistically compatible with no
sterile-neutrino oscillations. Again, we note that assuming Wilks' theorem we would still have
a $\approx 2\sigma$ hint for sterile neutrinos, whereas the MC analysis reduces this to
$\approx 1\sigma$.%

\begin{figure}[t]
	\centering
	\includegraphics[width=0.95\textwidth]{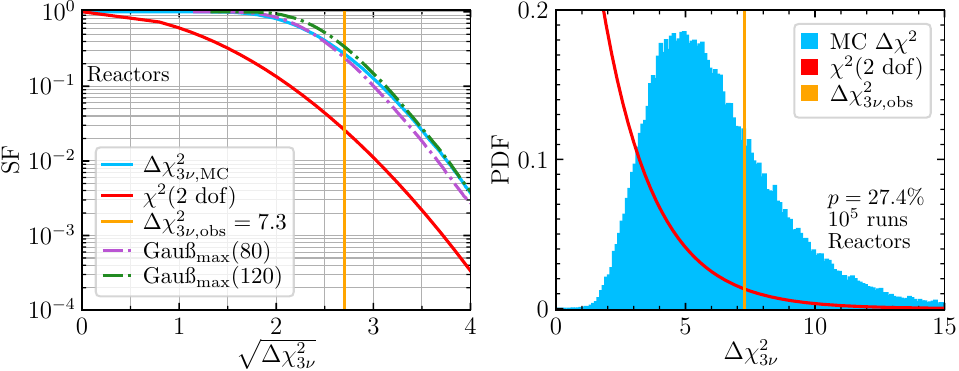}
	\caption{Survival function (SF, left) and probability
	distribution function (PDF, right) of \(\Delta\chi^2\) under the null hypothesis for the
	combination of all reactor experiments for \(10^5\) runs (blue). We also plot the \(\chi^2
	(\text{2 dof)}\) SF and PDF for comparison (red), as well as the maximum-Gauss
	distribution for $n=80$, 120, see ref.~\cite{Coloma:2020ajw}.
	}
	\label{fig:null-hyp-react}
\end{figure}%

This effect is further illustrated in \cref{fig:null-hyp-react}, where we show the probability
distribution function (PDF) and the survival function\footnote{The survival function is
defined as SF = $1 - $CDF, where CDF denotes the cumulative distribution function.} (SF) of
the $\Delta\chi^2_{3\nu}$ distribution for the combination of all reactor experiments,
obtained numerically by generating $10^5$ MC data sets. We see from the plot that the
distribution deviates considerably from the $\chi^2$ distribution with 2~dof. On the other
hand, in ref.~\cite{Coloma:2020ajw} it has been shown that for this type of oscillation
search, under certain idealizations, the SF for $\Delta\chi^2_{3\nu}$ follows the so-called
maximum-Gauss distribution, which is defined as the maximum value of $n$ independent
standard-normal variables. Therefore, in \cref {fig:null-hyp-react} we also compare the
numerical SF with this distribution for $n=80$ and 120 (here $n$ plays the role of an
effective number of degrees of freedom, and has been chosen in order to match the shape of the
SF). Note that small deviations from the maximum-Gauss distribution occur since some of the
assumptions under which it has been derived are not fulfilled exactly; for instance, due to
non-trivial systematic uncertainties, the data points in our fit are not all statistically
independent.%

\section{Gallium radioactive-source experiments}
\label{sec:gallium}

The gallium solar-neutrino detectors GALLEX and SAGE have been used to measure the neutrino
rate from radioactive $^{51}$Cr and $^{37}$Ar sources \cite{Hampel:1997fc,Kaether:2010ag,
Abdurashitov:1998ne,Abdurashitov:2005tb}, leading to results consistently lower than expected.
This so-called gallium anomaly could potentially find its explanation by sterile-neutrino
oscillations \cite{Acero:2007su,Kopp:2013vaa}. Recently, the BEST collaboration has performed
a dedicated source experiment using a $^{51}$Cr source embedded in a two-volume gallium
detector, confirming the previous hints at high significance~\cite{Barinov:2021asz}; see also
ref.~\cite{Barinov:2021mjj}.%

We summarize the gallium data in \cref{tab:gallium}, quoting the ratios of observed to
expected event rates.
\begin{table}[t]
	\setlength\tabcolsep{3pt}
	\centering
	\begin{tabular}{|c|c|c|c|c|c|}
		\hline
		\multirow{2}{*}{GALLEX \cite{Kaether:2010ag}}		& \multirow{2}{*}{GALLEX \cite{Kaether:2010ag}}		&
		\multirow{2}{*}{SAGE \cite{Abdurashitov:1998ne}	}	& \multirow{2}{*}{SAGE \cite{Abdurashitov:2005tb}}	&
		BEST \cite{Barinov:2021asz} & BEST\cite{Barinov:2021asz}  \cr
		 &  &
		 &  &
		(inner) & (outer) \\\hline
		$0.953 \pm 0.11$	& $0.812\pm0.10$	&
		$0.95\pm0.12$		& $0.791\pm0.084$	&
		$0.791\pm 0.044$	& $0.766\pm 0.045$ \\
		\hline
	\end{tabular}
	\caption{The ratio of observed to expected event rates for the gallium experiments. The
	quoted $1\sigma$ errors include statistical and uncorrelated experimental errors. The SAGE
	measurement \cite{Abdurashitov:2005tb} in the fourth column uses an $^{37}$Ar source; all
	other measurements are from a $^{51}$Cr source. The ratios are based on the Bahcall cross
	sections \cite{Bahcall:1997eg}.
	}
	\label{tab:gallium}
\end{table} %
The errors in the table include only statistical and uncorrelated
experimental errors. The errors for the two BEST data points are obtained by adding the statistical
error given in tab.~1 of ref.~\cite{Barinov:2021asz} and the uncorrelated experimental
systematic error of 2.6\% in quadrature. We observe that the reported ratios deviate from
unity by about 20\%, confirmed by BEST at high significance.%

\begin{table}[t]
	\centering
	\setlength\tabcolsep{1ex}
	\begin{tabular}{|l|cc|cc|}
		\hline
		Reference &
		$\sigma$(Cr)& $\sigma_{\rm g.s.}$(Cr)& $\sigma$(Ar) & $\sigma_{\rm g.s.}$(Ar) \\
		\hline
		Bahcall \cite{Bahcall:1997eg}	&
		$58.1 \pm 2.1$	&$55.2$			& $70.0\pm4.9$		& $66.2$		\\
		Kostensalo \emph{et al.}~\cite{Kostensalo:2019vmv}
		& $56.7\pm 1.0$	&$55.3\pm0.7$	& $68.0\pm1.2$		& $66.2\pm0.9$	\\
		Semenov \cite{Semenov:2020xea}	&
		$59.38\pm1.16$	&$55.39\pm0.19$	& $71.69\pm 1.47$	& $66.25\pm0.23$\\
		\hline
	\end{tabular}
	\caption{Cross section for $\nu_e$ detection on gallium, weighted by the energy lines
	for neutrinos produced from
	a Cr and an Ar source, for our three representative cross-section models. Units are
	$10^{-46}\,{\rm cm}^2$. We give the best estimate for the total cross sections as well as
	the cross section corresponding to the ground state (g.s.) transition only. For the
	Bahcall numbers we quote the (larger) upper error (see text); Bahcall does not provide an
	error estimate on $\sigma_{\rm g.s.}$.
	}
	\label{tab:galliumCS}
\end{table}%

An important ingredient in the interpretation of these results is the $\nu_e$-capture cross
section on gallium. The ratios in \cref{tab:gallium} are based on the traditional Bahcall 1997
cross sections~\cite{Bahcall:1997eg}, but for alternative cross sections they
have to be rescaled accordingly. Here we have considered two recent
re-evaluations of the cross sections from Kostensalo \emph{et al.\ }\cite{Kostensalo:2019vmv} and
Semenov \cite{Semenov:2020xea}, summarized in \cref{tab:galliumCS}. Note that in
\cref{tab:galliumCS} we give the best estimates for the total cross section for the
$\nu_e$-induced $^{71}{\rm Ga} \to ^{71}$Ge transition, which receives contributions from the
ground-state-to-ground-state (g.s.) transition as well as from transitions into excited states
of $^{71}$Ge. Two remarks should be made at this point. First, the matrix element of the g.s.
transition can be directly inferred from the electron-capture lifetime of $^{71}$Ge and the
corresponding phase-space factor. Second, the excited-state contributions add incoherently and
thus can only increase the cross section, that is, the g.s.\ transition provides a \emph{lower
bound} on the total cross section. Finally, note that the relatively large error in the
g.s.-only cross section of Kostensalo \emph{et al.\ }arises because it is based on their
shell-model calculation and not on the observed electron-capture parameters. Therefore, the
choice of this calculation as default will produce the most conservative result for the
g.s.-only analysis. At the same time, this choice is not fully consistent, since the
measured ground state strength is used to tune $g_A$ for the shell model calculation of
Kostensalo \emph{et al.}. Semenov's ground state cross section and uncertainty would
be a more consistent choice, but would lead to a much larger significance for the
gallium anomaly (see below).

In order to analyze gallium data, we define the following $\chi^2$ function:
\begin{align}
	\chi^2_{\rm Gallium} = \min_\xi\left[\sum_{i=1}^6 \frac{(R_i - (1+\xi) P_i)^2}{\sigma_i^2} +
	\frac{\xi^2}{\sigma_\xi^2}\right] \,.
\end{align}
Here, $R_i$ and $\sigma_i$ are the observed ratios and their uncertainties from
\cref{tab:gallium}, rescaled according to the adopted cross section, and $P_i$ are the
oscillation probabilities, averaged over the detector volumes and weighted by the
neutrino-energy lines from the corresponding source. The pull parameter $\xi$ describes the
correlated uncertainty on the cross section.\footnote{We assume the uncertainties of
$\sigma$(Cr) and $\sigma$(Ar) to be fully correlated. The combined fit is dominated by BEST,
and the impact of the SAGE argon datum is small; therefore this assumption has little
impact on the result.} In order to set the uncertainty $\sigma_\xi$ we proceed as follows: for
a given cross-section model, we first minimize with respect to $\xi$, adopting the total
uncertainty given in \cref{tab:galliumCS}; we then check whether the resulting cross section
at the pull minimum is smaller than the ground-state contribution; if this is the case, then we
switch to the smaller uncertainty of $\sigma_{\rm g.s.}$, making the pull stiffer. In this
way, we take into account the asymmetries of cross-section uncertainties. An analogous procedure is adopted when generating random values for the pull parameter for the MC studies.
\begin{table}[t]
	\centering
	\setlength{\tabcolsep}{4pt}
	\begin{tabular}{|l|cc|cc|cccc|}
		\hline
		 & \multicolumn{2}{|c|}{GALLEX}  & \multicolumn{2}{|c|}{\multirow{2}{*}{BEST}} &
		\multicolumn{4}{|c|}{\multirow{2}{*}{All gallium combined} } \cr
		& \multicolumn{2}{|c|}{\& SAGE} &\multicolumn{2}{|c|}{} &
		\multicolumn{4}{|c|}{} \\
		\hline
		Cross section	& $\Delta\chi^2_{3\nu}$		& $\#\sigma^{(W)}$
		& $\Delta\chi^2_{3\nu}$		& $\#\sigma^{(W)}$
		& $\sin^22\theta_{\rm min}$	& $\Delta m^2_{\rm min}$
		& $\Delta\chi^2_{3\nu}$		& $\#\sigma^{(W)}$ \\
		\hline
		Bahcall \cite{Bahcall:1997eg}				&
		3.7 & 1.4 & 31.3 & 5.2 & 0.35 & \SI{1.3}{eV^2} & 31.7 & 5.3 \\
		Kostensalo \cite{Kostensalo:2019vmv}	&
		4.9 & 1.7 & 31.5 & 5.2 & 0.32 & \SI{1.3}{eV^2} & 32.9 & 5.4 \\
		Semenov \cite{Semenov:2020xea}				&
		9.4 & 2.6 & 42.4 & 6.2 & 0.39 & \SI{1.3}{eV^2} & 44.7 & 6.4 \\
		Ground state								&
		3.4 & 1.3 & 29.7 & 5.1 & 0.29 & \SI{1.3}{eV^2} & 31.5 & 5.3 \\
		\hline
	\end{tabular}
	\caption{Significance of the gallium radioactive-source measurements for the previous GALLEX
	and SAGE results combined, the recent BEST result, and the combination of all gallium data
	using different evaluations of the detection cross section. For each data set, we give the
	$\Delta\chi^2$ of the null hypothesis and the corresponding significance in terms of
	Gaussian standard deviations, which we evaluate using Wilks' theorem, {\em i.e.,} assuming
	that the \(\Delta\chi^2\) is distributed as \(\chi^2(2\text{ dof})\). For the combined
	data, we also give the location of the oscillation parameters at the global minimum. For the
	ground-state-only analysis (last row) we assume the Semenov g.s.\ value and uncertainty
	({\em c.f.} \cref{tab:galliumCS}).}
	\label{tab:gallium-results}
\end{table}

In \cref{tab:gallium-results}, we report the results of our analysis of gallium data under the
sterile-neutrino-oscillation hypothesis adopting different assumptions about the detection
cross section. We give the value of $\Delta\chi^2_{3\nu}$ as defined in \cref{eq:T} and the
corresponding significance in units of Gaussian standard deviations. While previous results
from GALLEX and SAGE show only a weak hint for sterile-neutrino oscillations, the significance
of the BEST result is $>5\sigma$, independent of the assumption on the cross section. Note
that even the conservative ground-state-only analysis leads to more than $5\sigma$
significance. Let us stress that in order to calculate the significances in
\cref{tab:gallium-results}, we assume that $\Delta\chi^2_{3\nu}$ is distributed as $\chi^2(2
\text{ dof})$. For the combined analysis and the Kostensalo \emph{et al.\ }cross section we
have tested this assumption with a high-statistics MC study (\(10^9\) pseudo experiments). We
find that the \(\Delta\chi^2_{3\nu}\) distribution is more similar to \(\chi^2(1\text{ dof})
\). This reduction of the dof is related to the presence of the physical boundary $\sin^22\theta\ge0$, see \emph{e.g.,} ref.~\cite{Elevant:2015ska} for a discussion of this effect . The null hypothesis, when evaluated by Monte Carlo, yields
\begin{equation}
p_0=\num{2.7e-8} \,(5.6 \sigma) \,,\qquad\text{(gallium data, Kostensalo}\ et\ al.\ \text{cross section)}\,
\end{equation}
which is slightly higher than the corresponding value for a $\chi^2$ with 2 dof
(\(5.4\sigma\), see \cref{tab:gallium-results}) and more similar to the value for 1 dof, which
is \(5.7\sigma\).

\begin{figure}[t]
	\centering
		\centering
		\includegraphics[width=0.9\textwidth]{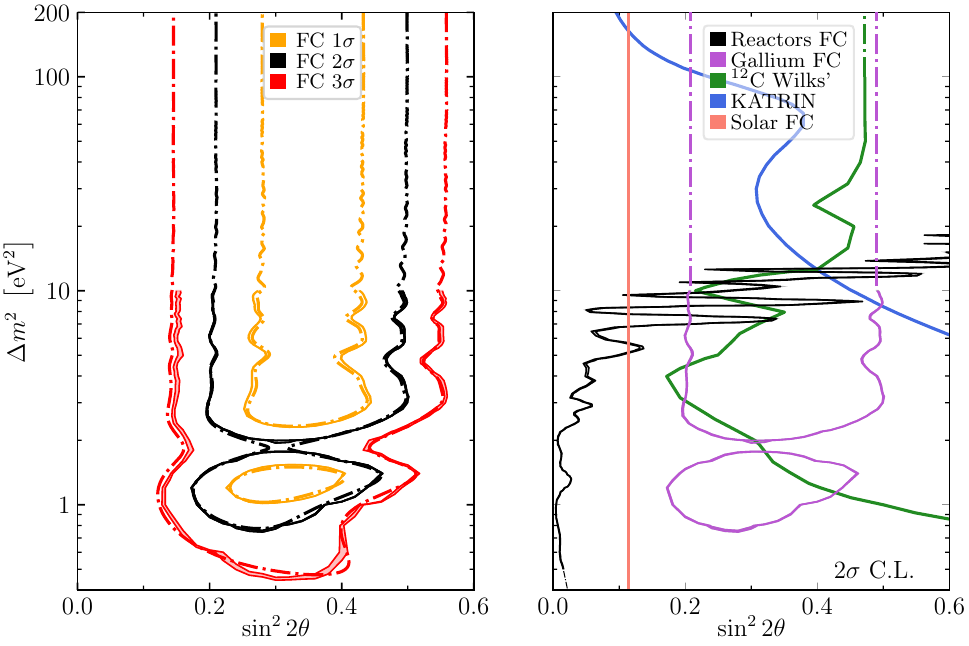}
		\caption{Left: confidence regions at 68.27\%, 95.45\%, 99.73\%~CL (1, 2, 3$\sigma$,
		respectively) for the Feldman-Cousins analysis (bands), where the bands indicate the 99\%
		confidence spread of the contour for \(10^5\) runs due to the finite size of our MC sample
		(see \cref{app:FC-error}). Dash-dotted curves are obtained under the assumption of Wilks'
		theorem (2 dof). Right: FC confidence regions for gallium, reactor, and solar data at
		2\(\sigma\). We superimpose the 95\% exclusion limit from the KATRIN collaboration
		\cite{KATRIN:2020dpx} (this curve fixes the lightest neutrino mass to zero) and the Wilks'
		95\% exclusion limit (2 dof) from \(\nu_{\rm e}\)-\(^{12}{\rm C}\) scattering from the
		LSND and KARMEN experiments \cite{LSND:2001fbw,Armbruster:1998uk}, taken from
		ref.~\cite{Kopp:2013vaa}. Dash-dotted lines are extrapolations assuming constant
		sensitivity in the mixing.
		}
		\label{fig:gallium}
\end{figure}%

In the following, we will adopt the Kostensalo \emph{et al.}~\cite{Kostensalo:2019vmv} cross
section as our default analysis for gallium data. In \cref{fig:gallium} (left panel), we show
the allowed regions in the sterile-neutrino parameter space for gallium data. An explanation
of these measurements in terms of sterile-neutrino oscillations requires rather large mixing
angles, $0.21\le \sin^22\theta \le 0.47$ at $2\sigma$. We observe that in this case the FC
confidence regions are in excellent agreement with $\Delta\chi^2$ contours under Wilks'
theorem assuming 2 dof. Many of the arguments leading to deviations from Wilks' theorem in the
reactor case \cite{Coloma:2020ajw} do not apply here. In particular, the effect is basically a
constant suppression of the rate, without oscillatory behaviour. Since the allowed region appears far away from the physical boundary (which dominates the distribution under the null-hypothesis, see above), we recover here the full freedom corresponding to the 2 parameters $\sin^22\theta$ and $\Delta m^2$.

In the right panel of \cref{fig:gallium} we compare the region preferred by gallium with
constraints from reactor data, solar neutrinos, the KATRIN experiment, and data on \(\nu_{\rm
e}\)-\(^{12}{\rm C}\) scattering from the LSND and KARMEN experiments. In the next section,
we are going to investigate in detail the compatibility of gallium, reactor and solar neutrino
data. The KATRIN constraint shown in \cref{fig:gallium} is taken from \cite{KATRIN:2020dpx},
where this curve has been derived by fixing the lightest neutrino mass to zero. This
constraint cuts off the allowed parameter space at large values of $\Delta m^2\gtrsim
100$~ev$^2$, but becomes quickly weaker for smaller mass-squared differences. We note that the
KATRIN limit would be further relaxed if the assumption of neglecting the light neutrino
masses is dropped \cite{KATRIN:2020dpx,Giunti:2019fcj}. The constraint from $\nu_e$ scattering
on $^{12}$C from LSND~\cite{LSND:2001fbw} and KARMEN~\cite{Armbruster:1998uk} is taken from
the re-analysis performed in ref.~\cite{Kopp:2013vaa}. This limit is in tension (at $\sim
2\sigma$ under Wilks' theorem) with the gallium region for $\Delta m^2 \lesssim 10$~eV$^2$.
However, note that the results from LSND/KARMEN and KATRIN
shown here were obtained assuming Wilks' theorem is applicable, which may lead to an
overestimation of the exclusion regions, similar to the data sets studied here. Thus, due to
the limited statistical power of the KATRIN and LSND/KARMEN constraints in the region of
interest, we focus below on gallium, reactor, and solar data.

\section{Global Fit Results and Consistency Tests} \label{sec:results-global}

In this section, we combine the reactor and gallium data discussed in \cref
{sec:reactor,sec:gallium}, respectively, and study their consistency. In addition, we take
into account information from solar neutrinos, which provides an important constraint on large
mixing angles. The solar-neutrino analysis adopted here is based on the simplified $\chi^2$
construction from ref.~\cite{Goldhagen:2021kxe}, which offers an efficient way to include
solar neutrino data in a MC study. We refer the interested reader to that reference for
details; a brief summary is provided in \cref{app:solar}.%

\Cref{tab:best-fit} shows the value of the $\chi^2$ per dof for the combination of reactor data, and the combination of reactor + solar, and reactor + gallium. The mild differences between the values of the $\chi^2_\text{min}$ in the last three rows of the table indicate that the usual goodness-of-fit test would not provide any insights concerning the mutual consistency of the different data sets, given the very large number of dof at hand. Therefore, in order to quantify the consistency between different data sets, we use the parameter
goodness-of-fit (PG) test \cite{Maltoni:2002xd,Maltoni:2003cu}, based on the test statistic
\begin{equation}
	\chi^2_{\rm PG} = \chi^2_{\rm min,comb} - \sum_k \chi^2_{{\rm min},k} \,,
\end{equation}
where the index $k$ labels the data sets to be tested for consistency. The first term on the
right-hand side corresponds to the global $\chi^2$ minimum, whereas the second term is the sum
of the $\chi^2$ minima of each data set individually. Under Wilks' theorem, $\chi^2_{\rm PG}$
will follow a $\chi^2$ distribution with $\sum_k P_k - P$ dof, where $P_k$ is the number of
parameters, on which the data set $k$ depends, and $P$ is the total number of parameters of
the model \cite{Maltoni:2002xd,Maltoni:2003cu}. For our cases of interest, we have $P_{\rm
reactor}= P_{\rm gallium} = P = 2$, corresponding to $\sin^22\theta$ and $\Delta m^2$, while
$P_{\rm solar} = 1$ --- in the limit relevant for us here, solar data are independent of
$\Delta m^2$ and depend only on $\sin^22\theta$.

\begin{figure}[t]
		\centering
		\includegraphics[width=0.6\textwidth]{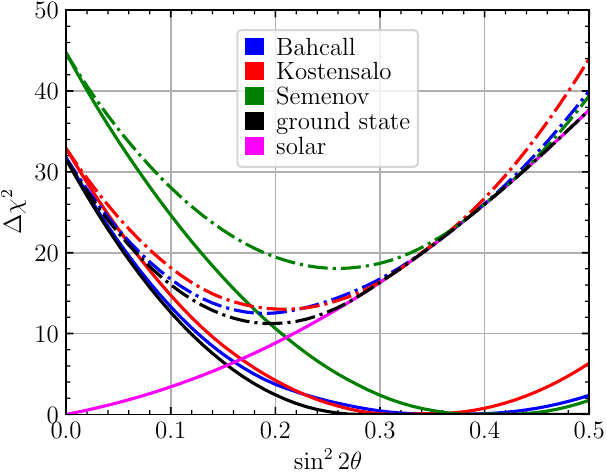}
		\caption{Solid: $\Delta\chi^2$ profiles (marginalizing $\Delta m^2$) for combined
		gallium data for four different assumptions on the detection cross section, see
		\cref{tab:galliumCS}. The magenta curve corresponds to solar neutrino data, and the
		dashed curves show the sum $\Delta\chi^2_{\rm gallium} + \Delta\chi^2_{\rm solar}$}
		\label{fig:gallium-vs-solar}
\end{figure}%
Let us first investigate the consistency of the solar and gallium data sets.
\Cref{fig:gallium-vs-solar} shows the $\Delta\chi^2$ profiles for gallium data under various
assumptions of the detection cross section, compared to the one from solar data. The
dash-dotted curves correspond to $\Delta\chi^2_{\rm gallium}+\Delta\chi^2 _{\rm solar}$; the
value of $\chi^2_{\rm PG}$ for the consistency of solar and gallium data is given by the
minimum of the dashed curves. We see that the Bahcall, Kostensalo and g.s.\ cross sections
all give very similar results for $\chi^2_{\rm PG}$. Even under the ground-state-only
assumption, we would obtain $\chi^2_{\rm PG} = 11$, corresponding to $3.3\sigma$ tension. For
the Semenov cross section, we find $\chi^2_{\rm PG} = 18$ and the tension is at the
$4.2\sigma$ level. Also, note that here we use the GS98 solar model \cite{Vinyoles:2016djt},
which leads to conservative limits. If the AGSS09 model \cite{Vinyoles:2016djt} were used
instead, then the solar constraint would get significantly stronger \cite{Goldhagen:2021kxe},
making the tension with gallium even more severe.%

\begin{table}[tb!]
	\setlength\tabcolsep{1ex}
	\centering
	\begin{tabular}{|l|c|cc|cc|}
		\hline
		Data set & $\chi^2_{\rm PG}$/dof & $p^{(W)}$ & $\#\sigma^{(W)}$ &
		$p_\text{b.f.}$ & $\#\sigma_\text{b.f.}$ \\
		\hline
		Reactor	vs	Solar			& 0.65/1	& 0.42	& 0.8	 & 0.39 & 0.9 \\
		Reactor vs Gallium		& 1.4/2		& 0.50	& 0.67 & 0.62 & 0.5	\\
		Solar vs Gallium			& 13.0/1	& \num{3.1e-4}	& 3.6	& \num{1.6e-3} & 3.2 \\
		Reactor vs Solar vs Gallium	& 15.6/3	& \num{1.4e-3}	& 3.2 & \num{5.1e-3} & 2.8	\\
		\hline
	\end{tabular}
	\caption{Consistency test of the various data sets based on the parameter goodness-of-fit
	\cite{Maltoni:2002xd,Maltoni:2003cu}. In the middle columns, the $p$-values and number of
	Gaussian standard deviations are evaluated under the assumption of Wilks' theorem. The final
	two columns show results derived from MC simulations, generating pseudo-data fluctuations
	around the best-fit (b.f.) prediction.
	\label{tab:consistency}}
\end{table}%

Next, let us study the consistency when reactor data are also considered. The results of
various consistency checks are reported in \cref{tab:consistency}, where, for concreteness, we
have adopted the Kostensalo \emph{et al.}~cross section for the gallium experiments. The last
two columns show our results for the PG test, where the distribution of \(\chi^2_{\rm PG}\) is
calculated using MC simulations. In doing so, the pseudo-data fluctuations are generated by
using the prediction for the corresponding best-fit parameters as the pseudo-data mean. We
find that both the reactor and solar data, as well as reactor and gallium data, are fully
compatible. For comparison, the middle columns in \cref{tab:consistency} show the expected
results obtained under the assumption of Wilks' theorem. Although the differences in $p$-value
with respect to the MC result are small in the case of consistent data sets, we find a larger
discrepancy in the case of inconsistent data sets. This is shown in the last two rows of the
table, where our MC simulations show a reduction of $\sim 0.5\sigma$ with respect to the
expected result under the assumption that Wilks' theorem holds.

\begin{figure}[t]
	\centering
	\includegraphics[width=0.9\textwidth]{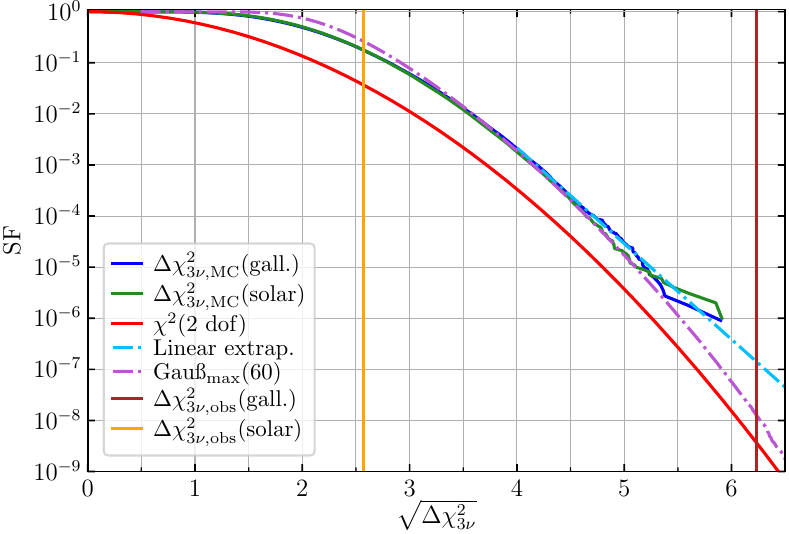}
	\caption{Survival function of \(\sqrt{\Delta\chi^2_{3\nu}}\) under the null hypothesis for
	reactor data combined with solar (green) or with gallium (blue), for \(10^6\) runs.
	For comparison we also show
	the \(\chi^2(\text{2 dof)}\) SF (solid red line), the
	maximum-Gauss distribution for $n=60$ (dash-dotted purple), see \cite{Coloma:2020ajw}, and a
	linear extrapolation of the tail of the gallium SF (dash-dotted light blue).
	}
	\label{fig:null-hyp}
\end{figure}%

Given the strong tension between solar and gallium data, we will not combine these two data
sets in the following but consider only the combinations reactor+solar and reactor+gallium
separately. The best-fit points and \(p\)-values for the null hypothesis of these
combinations can be found in the last rows of \cref{tab:best-fit}. While the combination of
solar and reactor data shows no significant hint for sterile-neutrino oscillations, the
reactor+gallium combination prefers oscillations over the null hypothesis with $\Delta\chi^2_
{3\nu} = 38.8$. Assuming Wilks' theorem (2~dof), this would correspond to $p_0^{(W)}=3.8\times
10^{-9}$ ($5.9\sigma$). Such a high significance makes it impossible to calculate a $p$-value
by MC.

In \cref{fig:null-hyp}, we show the MC distribution of the test statistic for the null
hypothesis, $\Delta \chi^2_{3\nu}$, defined in \cref{eq:T}. Both the reactor+solar and
reactor+gallium datasets have an almost identical distribution which differs significantly
from a $\chi^2$ distribution, but has a similar shape as the maximum-Gauss distribution
discussed in \cite{Coloma:2020ajw}. This suggests that in both cases the statistical
properties are dominated by the reactor data set. We can use this observation to extrapolate
the SF of $\Delta\chi^2_{3\nu}$ to the observed value for reactor+gallium. Assuming a
maximum-Gauss distribution with $n=60$, which gives a good fit to the MC distribution in the
region where it can be simulated (\emph{c.f.,} purple dash-dotted curve in
\cref{fig:null-hyp}), we find %

\begin{equation}
p_0=\num{1.4e-8} \,(5.7 \sigma) \,,\qquad\text{(reactor+gallium, max.-Gauss extrapolation)}\,.
\end{equation}
If we instead use a linear extrapolation of the tail in the log-plot, (light-blue dash-dotted line in \cref{fig:null-hyp}) we would obtain
\begin{equation}
p_0=\num{1.4e-7} \,(5.3 \sigma) \,,\qquad\text{(reactor+gallium, linear extrapolation)}\,.
\end{equation}
In any case we confirm the $\gtrsim 5\sigma$ indication in favor of sterile-neutrino
oscillations of combined reactor+gallium data, which is of course driven by the BEST result.
Under the adopted assumptions for the BEST analysis, this is a statistically robust result.

\begin{figure}[t]
	\centering
	\includegraphics[width=\textwidth]{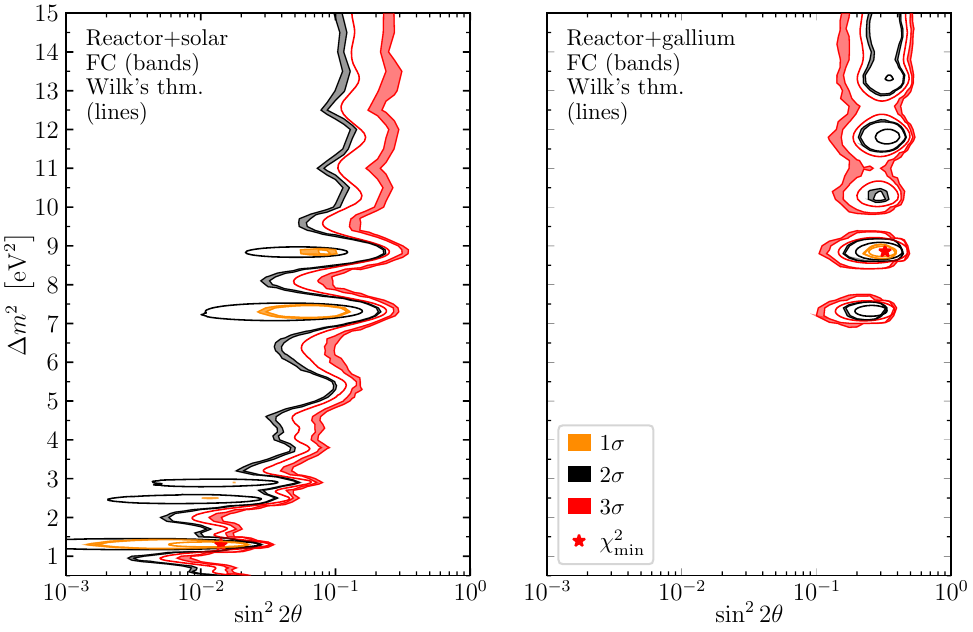}
	\caption{Confidence regions at 68.27\%, 95.45\%, 99.73\%~CL (1, 2, 3$\sigma$, respectively)
	for reactor+solar data (left), and reactor+gallium data (right). The bands correspond to
	the Feldman-Cousins analysis, where the width of the bands indicates the 99\% confidence
	spread of the contour for \(10^4\) runs due to the finite size of our MC sample (see
	\cref{app:FC-error}). Thin curves are obtained under the assumption of Wilks' theorem.}
	\label{fig:FCMC-all}
\end{figure} %

In \cref{fig:FCMC-all} we show the allowed regions for reactor+solar data (left) and reactor
+ gallium data (right). For the reactor+solar data we find no closed contours at $2\sigma$
for the FC analysis, consistent with the null-hypothesis being allowed at $1.3\sigma$. The
main effect of solar neutrino data is to exclude the large mixing angles in the region $\Delta
m^2 \gtrsim 10$. For the reactor+gallium analysis the determination of the mixing angle
$0.11<\sin^22\theta<0.53$ ($3\sigma$) is dominated by gallium data, whereas $\Delta m^2$ is
determined by Neutrino-4 reactor data. The best-fit point is located at
\begin{align}
	\sin^22\theta=0.32\,,\quad \Delta m^2=8.86 \quad\text{(best fit, reactor+gallium data)}\,,
\end{align}
however, several local minima are present at $2\sigma$. In both panels we note again that the FC confidence regions are larger by roughly $1\sigma$ compared to contours based on Wilks' theorem.

\section{Summary and Conclusions}
\label{sec:conclusions}

We have performed a global Feldman-Cousins analysis of the reactor short-baseline experiments
DANSS, NEOS, PROSPECT, STEREO and Neutrino-4, past gallium data from SAGE and GALLEX and the
recent result of the BEST experiment, as well as solar neutrino data. The analysis is
performed in a 3+1-neutrino-oscillation framework. In general, our results show that a full
Feldman-Cousins treatment reduces all $p$-values by about $1\sigma$ as compared to a naïve
application of Wilks' theorem, as shown in \cref{tab:best-fit}. Our main findings are that no
reactor experiment individually favors oscillation at more than $2.4\,\sigma$, with the
strongest hint coming from Neutrino-4. Combined reactor data are consistent with no
sterile-neutrino oscillations at $1.1\,\sigma$, and solar neutrino data are fully compatible
with this result. On the other hand, gallium data favor a rate deficit at more than
$5\,\sigma$, even under conservative choices for the cross-section uncertainty. This result is
compatible with reactor data, and would then point to values of $\Delta m^2\simeq\numrange
7{12}\,\si{eV^2}$, driven by the Neutrino-4 result. At the same time, the gallium and solar
results are in tension, according to a parameter goodness-of-fit test~\cite {Maltoni:2003cu}
by more than $3\,\sigma$. One important caveat to our findings is that we take each
collaboration's statements about systematic uncertainties at face value.

In comparison to earlier global fits of reactor and gallium data, for instance
refs.~\cite{Dentler:2017tkw,Gariazzo:2018mwd,Berryman:2020agd}, we find that the newest data
from DANSS, NEOS, PROSPECT and STEREO are consistent with the no-oscillation hypothesis; the
evidence in favor of oscillation stems entirely from the latest BEST gallium data, consistent
with hints at much weaker significance from previous gallium measurements and the Neutrino-4
reactor data, which all point to larger mixing angles and mass-squared splittings. These data
sets are in mutual agreement, with a parameter goodness-of-fit $p$-value of 0.5. The reactor
experiments not seeing any evidence for oscillation are simply insensitive to the region
preferred by Neutrino-4 and the gallium data. Although the Neutrino-4 results have been
challenged on experimental grounds~\cite{PROSPECT:2020raz,Danilov:2018dme,Danilov:2020rax,Giunti:2021iti}, we take the
results published by the collaboration as the basis for our analysis. Neutrino-4 alone prefers
oscillation at $2.4\,\sigma$ (with a $p$-value of $1.5\%$), whereas the gallium data exclude
the no-oscillation hypothesis at more than $5\,\sigma$. For gallium data, the main systematic
is the cross section for the reaction $^{71}{\rm Ga}\,+\,\nu_e\rightarrow {^{71}{\rm Ge}}\,+\,
e^-$; we have performed a literature survey of several cross-section calculations. The
difficulty arises from excited-state contributions, since the ground-state-to-ground–state
transition matrix element can be directly determined from the observation of the electron
capture of $^{71}$Ge. However, the excited-state contributions add incoherently and thus can
only increase the cross section, making the effect larger. Even considering only the
ground-state contribution to the cross section, we find the result is $>5\,\sigma$
significant, see \cref{tab:gallium-results}. If this result is due to some systematic effect,
then it seems unlikely that it is related to the cross section. Note that the results from the
two detector volumes in BEST are practically identical and thus this result (as was the case
for past gallium results) is an absolute rate measurement, leaving room for some common,
as-yet-unidentified, systematic affecting the overall normalization of the event rates.

On the other hand, solar data strongly reject sterile-neutrino oscillations with large mixing
angles for the whole range of mass-squared splittings relevant here. As a consequence, solar
data are in significant tension with the gallium data (with a parameter goodness-of-fit of at
least $3\,\sigma$, see \cref{tab:consistency,fig:gallium-vs-solar}). The evidence in favor of
oscillation from reactors is much weaker and hence there is no tension in a global fit between
solar and reactor data. The solar bound presumably could be evaded by constructing a model in
which sterile neutrinos experience some anomalous matter effect in the Sun, but this goes
beyond the 3+1 model considered here.

This leaves the question of how to test our findings. In the context of 3+1 oscillations, note
that the oscillation length of a 4\,MeV neutrino with $\Delta m^2=10\,\mathrm{eV}^2$ is only
about 0.5\,m, making the direct observation of oscillations in a reactor experiment very
challenging. The KATRIN experiment is, in principle, sensitive to sterile neutrinos down to
$\Delta m^2\sim 10\,\mathrm{eV}^2$, but a strong exclusion limit can be only obtained if a
prior is put on the effective parameter $m_{\nu_e,\text{eff}}^2>0$. This prior is well
motivated by physics, but systematic effects in beta-decay endpoint experiments have
consistently led to measured values of $m_{\nu_e,\text{eff}}^2$ being negative (for a recent
review see {\em e.g.,} ref.~\cite{Formaggio:2021nfz}). Given the strong tension with solar
data and the fact that large values of $\Delta m^2\sim10\,\mathrm{eV}^2$ are also in tension
with standard cosmology, the notion that we are dealing with new physics different from a
vanilla sterile neutrino is worthwhile to investigate, as is the search for
as-yet-unidentified experimental systematics.

\begin{acknowledgments}
	We thank Joachim Kopp and Pedro Machado for providing us the results of their $^{12}$C LSND
	and KARMEN analysis. J.M.B. acknowledges support from the National Science Foundation,
	Grant PHY-1630782, and the Heising-Simons Foundation, Grant 2017-228. J.M.B. further thanks
	the Network for Neutrinos, Nuclear Astrophysics, and Symmetries
	(\href{https://n3as.berkeley.edu}{N3AS}) for encouragement and support; the University of
	Kentucky, where much of this work was completed; and the Institute for Nuclear Theory at the
	University of Washington for its kind hospitality and stimulating research environment. This
	research was supported in part by the INT's U.S. Department of Energy grant No.
	DE-FG02-00ER41132. Albert Zhou thanks the Doctoral School ``Karlsruhe School of Elementary
	and Astroparticle Physics: Science and Technology (KSETA)'' for financial support through
	the GSSP program of the German Academic Exchange Service (DAAD). The work of PC is supported
	by Grant RYC2018-024240-I funded by MCIN/AEI/10.13039/501100011033 and by ``ESF Investing in your future''.
	This project has received support from the European Union’s Horizon 2020 research and
	innovation program under the Marie Sklodowska-Curie grant agreement No 860881-HIDDeN. The
	authors acknowledge use of the HPC facilities at the IFT (Hydra cluster), and support of the
	Spanish Agencia Estatal de Investigaci\'on through the grant ``IFT Centro de Excelencia
	Severo Ochoa SEV-2016-0597'', as well as from Grant  PID2019-108892RB-I00 funded by MCIN/AEI/10.13039/501100011033 and by ``ERDF A way of making Europe''. The work of P.H. was supported by the U.S. Department of
	Energy Office of Science under award number DE-SC00018327.
\end{acknowledgments}

\appendix

\begin{subappendices}
	\renewcommand{\setthesubsection}{\Alph{section}.\arabic{subsection}}
	\section {Details on the Experiment Simulations and Data Analyses}
	\label{app:exp-impl}
	\setcounter{equation}{0}

	In this appendix, we describe our analyses of the reactor experiments, as well as our
	treatment of solar neutrino data. While our analyses of reactor experiments are insensitive
	to the modeling of the (differential) flux $\d\Phi/\d E_\nu$ and inverse-beta-decay cross
	section $\sigma_\text{IBD}$, we must make some choices for these in order to perform our
	simulations. When necessary, we simulate event rates using the Huber-Mueller spectra
	\cite{Huber:2011wv, Mueller:2011nm} and the Beacom-Vogel cross section \cite{Vogel:1999zy},
	unless noted otherwise. We reiterate that our results are insensitive to these choices.

	\subsection {DANSS} \label{app:DANSS}%

	The flux of neutrinos is determined by using the fuel fission fractions provided by DANSS
	\cite[p.\ 5]{DANSS-2021}. We assume the reconstructed energy \(E_\text{rec}\) is
	distributed as a Gaussian with mean \(E_\nu\) and width
	\begin{equation}
		\begin{split}
			\frac{\sigma(E_\nu)}{E_\nu} = 0.352 + 0.024\cdot (E_\nu/\mathrm{MeV}) -
			0.148\sqrt{E_\nu /\mathrm{MeV} } + \frac{0.040}{\sqrt{E_\nu/\mathrm{MeV}}},
		\end{split}
	\end{equation}
	encoded by the reconstruction matrix \(R(E_{\rm rec},E_\nu)
	\). The width is taken from ref.~\cite[fig.\ 5]{Danilov:2014vra} but increased by 10\% to
	better match the results of the collaboration.

	Because the baseline is so short, the finite size of the reactor core and detector needs
	to be taken into account. The detector is assumed to be a cube of \SI1{m^3} volume
	\cite{Alekseev:2016llm} and the reactor size is taken to be \SI{4.5}m. This number is set
	to be slightly higher than the value quoted in ref.~\cite{DANSS-2021} in order to obtain a
	better match to the results of the collaboration, and we assume that this is due to
	diagonally travelling neutrinos having a longer path length. Thus, for each baseline \(I
	\equiv B,M,T\) (bottom, middle, top) the predicted number of events is computed as
	\begin{equation} \label{eq:DANSS-pred}
		\begin{split}
			N_I^i(\Delta m^2) = N_{0,I}^i - \sin^22\theta \int_{L_I-0.5}^{L_I+0.5}\d x
			\int_{x-2.25}^{x+2.25}\d L \int_{E_i}^{E_{i+1}}\d E_\text{rec}
			\int_0^\infty\d E_\nu \\
			\bigg[ \sigma_\text{IBD}(E_\nu) \frac{d\Phi}{dE_\nu}(E_\nu) R(E_{\rm rec},E_\nu)
			\frac{\sin^2\frac{\Delta m^2 L}{4 E_\nu}}{L^2} \bigg].
		\end{split}
	\end{equation}

	The DANSS analysis is performed in terms of the ratios between the number of events at the
	different locations as outlined in \cref{sec:exp-reactor}, see \cref{eq:DANSS-R}. The
	predicted values for the ratios can be written in terms of the event rates in
	\cref{eq:DANSS-pred} as
	\begin{equation} \label{eq:DANSS-R-pred}
		\begin{aligned}
			P^{i}_1 = R_1^{3\nu} \frac{N^i_B}{N^i_T}, &&
			P^{i}_2 = R_2^{3\nu} \frac{N^i_M}{\sqrt{N^i_B N^i_T}} \, ,
		\end{aligned}
	\end{equation}
	where the no-oscillation predicted ratios \(R_1^{3\nu}, R_2^{3\nu}\) can be found in
	ref.~\cite[p.~17]{DANSS-2021}.

	A $2\times 2$ covariance matrix (with indices $a,b \in [1,2]$) is defined for each energy
	bin $i$:
	\begin{equation} \label{eq:DANSS-cov-matrix}
		V^i_{ab} = \sum_{I=B,M,T} \pfrac{R^i_a}{N^i_I}\pfrac{R^i_b}{N_I^i} \sigma_{N_I^i}^2 =
		\sum_{I=B,M,T} \pfrac{R_a^i}{n_I^i}\pfrac{R_b^i}{n_I^i}\frac{n_I^i}{\Delta t_I}
	\end{equation}
	where we have assumed that the uncertainty of the event rates in each bin is statistics
	dominated ($\sigma_{N_I^i} = \sqrt{N_I^i}$). This describes correlations due to the fact
	that \(P_{1,2}^i\) contain common factors of \(N_{B,T}^i\).

	The final ingredient in the computation of the $\chi^2$ function for DANSS is the effect of
	systematic uncertainties. We introduce two nuisance parameters \(\kappa_{1,2}\) for each
	of the ratios measured, associated with the relative efficiency, and add a pull term for
	each of them. The \(\chi^2\) function for DANSS is then defined as
	\begin{equation} \label{eq:DANSS-chisq-defn}
		\begin{split}
			\chi^2_{\rm DANSS}=\min_{\kappa_1, \kappa_2}
			\left\{\sum_{i=1}^{36} \sum_{a,b=1}^2
			[ D^i_a - (1 + \kappa_a) P^i_a] (V^i)^{-1}_{ab} [ D^i_b - (1 + \kappa_b) P^i_b]
			+ \frac{\kappa_1^2 + \kappa_2^2}{\sigma_{\rm{sys}}^2}\right\},
		\end{split}
	\end{equation}
	where \(\sigma_{\rm sys}=0.2\%\), and $D^i_a$ corresponds to the ratios measured by the
	experiment. The minimization is performed analytically, noting that the problem is to
	minimize a degree-two multinomial in the nuisance parameters.

	By following this procedure, we find good agreement between our \(\chi^2\) map and the
	result reported in ref.~\cite[p.\ 16]{DANSS-2021} by the collaboration. A comparison of
	the best-fit points is given in \cref{tab:best-fit}.

	To generate the pseudo data, we first compute the mean value for each bin as
	\begin{equation}
		\begin{aligned}
			\Braket{ n^i_T}=n_T^{i,\text{obs}}, \quad
			\Braket{ n^i_M}&=n_T^{i,\text{obs}}P^i_2 \sqrt{P^i_1}, \quad
			\Braket{ n^i_B}&=n_T^{i,\text{obs}}P^i_1 \, .
		\end{aligned}
	\end{equation}
	Statistical fluctuations are then injected by randomly sampling a Gaussian distribution
	for each bin, centered around its mean value and with standard deviation $\sigma_{I,i} =
	\sqrt{\Braket{n_I^i}/\Delta t_I}$.

	\subsection {NEOS} \label{app:NEOS}%

	The NEOS analysis uses the neutrino spectrum reported from the Daya Bay collaboration,
	obtained from ref.~\cite{DayaBay:2016ssb}. Due to this complication, and their unknown
	error response, we obtain our prediction by applying a ratio to the collaboration's
	no-oscillation prediction.

	Let us first denote the collaboration's prediction in the absence of oscillations by
	\(S_{i,0}\). We obtain \(S_{i,0}\) by dividing the ratios from ref.~\cite[fig.\ 3c]
	{NEOS:2016wee} through the extracted signal points (corresponding to the observed data
	minus the expected background) from ref.~\cite[fig.\ 3a]{NEOS:2016wee}. Doing this
	normalizes the sum $\sum_i S_{i,0} = 1891.6$.

	To obtain the prediction $S_i$ for the sterile-neutrino hypothesis, we then rescale\footnote
	{This rescaling is only approximate, due to the incomplete information on the detector
	response. However it should be accurate as long as the effects due to our mismodeling of
	the detector response are comparable for the numerator and the denominator.} the
	no-oscillation prediction by the ratio
	\begin{equation} \label{eq:neos-pred}
		S_i = S_{i,0} \frac{\DB_i (\sin^22\theta, \Delta m^2 )}{\DB_i (\sin^22\theta=0)} \, ,
	\end{equation}
	where $\DB_i$ is a function defined for each bin $i$ in reconstructed energy as
	\begin{equation} \label{eq:DB-4nu}
		\begin{split}
			\DB_i(\sin^22\theta, \Delta m^2) & =
			\left( 1 - \frac{\sin^22\theta}{2} \right)^{-1} \times \\
			&\int\d E_\nu \mathcal{R}_i(E_\nu)\times \left[\sigma_\text{IBD}\frac{\d\Phi}{\d E_\nu}\right](E_\nu)
			\times \Braket{ \frac{P_{ee}}{L^2} } (\sin^22\theta,\Delta m^2, E_\nu) \, ,
		\end{split}
	\end{equation}
	and
	\begin{equation} \label{eq:NEOS-Ri}
		\mathcal{R}_i(E_\nu) = \int_{E_i}^{E_{i+1}}
		R(E_{\rm rec},E_\nu) \d E_{\rm rec} \, .
	\end{equation}
	Here, $R(E_{\rm rec},E_\nu)$ stands for the reconstruction matrix (see below for an
	exact definition), while \(\left[\sigma_\text{IBD}\dderiv{\Phi}{E_\nu}\right](E_\nu)\) is
	the flux-weighted cross section in terms of the true neutrino energy, taken from
	ref.~\cite{DayaBay:2016ssb}. In \cref{eq:DB-4nu}, we have also defined a weighted
	oscillation probability with the inverse baseline squared as
	\begin{equation}
		\Braket{ \frac{P_{ee}}{L^2} } (\sin^22\theta,\Delta m^2, E_\nu) =
		\int_{L_\text{min}}^{L_\text{max}} P_{ee}
		\left( \sin^22\theta,\frac{\Delta m^2 L}{E_\nu} \right) \frac{\rho(L)}{L^2}\d L \,,
	\end{equation}
	where $P_{ee}$ is the 3+1-oscillation probability at NEOS; $L_{\rm min,max}$ are the
	minimum, maximum baselines respectively; and \(\rho(L)\) is the baseline distribution
	(extracted from ref.~\cite[p.\ 8]{Oh-NEOS:2017}).

	Note that the Daya Bay spectrum from ref.~\cite{DayaBay:2016ssb} unfolds the
	three-neutrino oscillations, and, since Daya Bay would be sensitive to both \(3\nu\) and
	\(4\nu\) oscillations, we need to remove distortions that would be introduced under a
	different hypothesis. For the large values of \(\Delta m^2\) considered in this work, we
	assume that the oscillations due to the new frequency, introduced by the sterile neutrino,
	average out at Daya Bay, which leads to the factor \([1-\sin^22\theta/2]^{-1}\) in
	\cref{eq:DB-4nu}. Corrections from different isotope fractions are applied as per
	ref.~\cite{NEOS:2016wee}.

	The reconstruction matrix $R(E_{\rm rec},E_\nu)$ has two components. For the first, we take
	the approximation from ref.~\cite{Huber:2016xis} for positron energy loss,
	\begin{equation} \label{eq:NEOS-Replus}
		\renewcommand{\arraystretch}{2.4}
		R_{e^+}(E_{\rm rec}) = \left\{
			\begin{array}{ll}
				Z + \left( \dfrac{1}{\sqrt{2\pi}\sigma(E_\text{avg})} - Z \right) \exp
				\left\{-\dfrac{(E_\text{rec}-E_\text{avg})^2}{2\sigma^2(E_\text{avg})}\right\}
				& \text{for} \;\; E_\text{rec} < E_\text{avg} \, , \\
				\dfrac{1}{\sqrt{2\pi}\sigma(E_\text{avg})} \exp
				\left\{-\dfrac{(E_\text{rec}-E_\text{avg})^2}{2\sigma^2(E_\text{avg})}\right\}
				& \text{for} \; \; E_\text{rec} > E_\text{avg} \, ,
			\end{array} \right.
	\end{equation}
	where \(Z=0.01\). Here, the energy resolution is parametrized as~\cite[fig.\ 3]{Ko:2018sx}
	\begin{equation}
		\sigma(E) = \left[ 2 \cdot 10^{-2} + 1.2\cdot 10^{-3} \frac{E}{\mathrm{MeV}} +
		0.2\cdot 10^{-3} \left(\frac{E}{\mathrm{MeV}}\right)^2 \right]^{1/2} \, ,
	\end{equation}
	and the average energy \(E_\text{avg}\) includes the non-linear detector response,
	\begin{equation}
		E_\text{avg} = 1.01\cdot E_p f(0.41\cdot E_p - 0.05),
	\end{equation}
	where \(f(\,\cdot\,)\) is taken from the supplementary material of
	ref.~\cite{DayaBay:2016ggj} and \(E_p = E_\nu - \Delta_{np} + m_e\), with $m_e$ being the
	electron mass and \(\Delta_{np}\) the neutron-proton mass difference. The choice of
	parametrization for the function $f$ was simply made by tuning \(AE_pf(aE_p+b)\) to the
	results of the collaboration, as no detector response was provided. Besides \cref
	{eq:NEOS-Replus} we also consider a second contribution from escaping 511-\si{keV}
	annihilation photons, with a relative height \(0.5\) as per ref.~\cite[p.\ 15]
	{Oh-NEOS:2017}:
	\begin{equation} \label{eq:NEOS-Rgamma}
		R_\gamma (E_{\rm rec}) = 0.5 \times \frac{1}{\sqrt{2\pi}\sigma(E_\text{avg})} \exp
		\left\{-\frac{(E_\text{rec}+m_e-E_\text{avg})^2}{2\sigma^2(E_\text{avg})}\right\} \, .
	\end{equation}
	The final reconstruction matrix $R(E_{\rm rec},E_\nu)$ is computed by adding the
	contributions from \cref{eq:NEOS-Replus,eq:NEOS-Rgamma}. However, for the last bin we find
	that this approach does not provide good agreement with ref.~\cite{NEOS:2016wee}. Thus,
	the entry in this last bin is estimated as a piecewise constant (see the inset of \cite
	[fig.\ 3a]{NEOS:2016wee}) so that the data/Daya Bay ratio is correctly reproduced.
	Finally, once the reconstruction matrix has been computed, the integral over reconstructed
	energy in \cref{eq:NEOS-Ri} is done analytically.

	Following ref.~\cite{NEOS:2016wee}, we define the NEOS covariance matrix as \(V_{ij} =
	V^\text{syst}_{ij} + V^\text{stat}_{ij}\). The matrix $V^\text{stat}$ accounts for the
	statistical errors and is defined as
	\begin{equation}
		\label{eq:NEOS-V-stat}
		V^\text{stat}_{ij} = \delta_{ij} \left[\frac{S_i^\text{obs} + B_i}{\Delta t_\text{on}} +
		\frac{B_i}{\Delta t_\text{off}}\right] \, ,
	\end{equation}
	where $B_i$ corresponds to the background events and $S_i^\text{obs}$ corresponds to the
	observed background-subtracted events per bin $i$ (that is, the black points in \cite[fig.\
	3a]{NEOS:2016wee}), while $\Delta t_{\rm on}$ and $\Delta t_{\rm off}$ correspond to the
	data-taking periods in days, with the reactor on and off, respectively. In
	\cref{eq:NEOS-V-stat} the first term is the Poisson error, while the second one stems from
	the fact that the background rate was measured during the (relatively short) reactor-off
	time period and therefore is not known precisely. On the other hand, the contribution
	$V^\text{sys}$ is defined as
	\begin{equation}
		\label{eq:NEOS-V-syst}
		V^\text{syst}_{ij} = \mathcal N^2\iint\d E_\nu\d E_\nu'\,V^\text{DB}(E_\nu,E_\nu')
		\mathcal{R}_i(E_\nu) \mathcal{R}_j(E'_\nu) \,
	\end{equation}
	where \(V^\text{DB}\) is constructed through a bicubic interpolation of the covariance
	matrix in ref.~\cite[table 13]{DayaBay:2016ssb}, and the normalization constant
	$\mathcal{N}$ is determined as
	\begin{equation}
		\mathcal N = \frac{\sum_i S_{i,0}}{\sum_i\DB_i(0)} \int_{L_\text{min}}^{L_\text{max}}
		\frac{\rho(L)}{L^2}\d L \, .
	\end{equation}

	As for systematic uncertainties, we include a nuisance parameter for the overall flux
	normalization, $\xi$, but we add no penalty term to the $\chi^2$: it is left free in the
	fit. Thus, the NEOS $\chi^2$ function reads:
	\begin{equation} \label{eq:NEOS-chisq-defn}
		\chi^2_{\rm NEOS} = \min_\xi \sum_{i,j} \;(S_i^\text{obs} - \xi S_i) V_{ij}^{-1} (S_j^\text{obs} - \xi S_j),
	\end{equation}
	where the minimization over $\xi$ is done analytically.

	We compared our Bayesian exclusion curve (computed as in ref.~\cite[eq.\ (4)]
	{NEOS:2016wee}) to the result of the collaboration; in this method, the unnormalized
	probability distribution for a fixed \(\Delta m^2\) is \(\exp(-\Delta\chi^2/2)\). Our
	curve has extremely good agreement in the range \numrange12 \si{eV^2}.

	\begin{table}
		\centering
		\begin{tabular}{|c|cccc|}
			\hline
			\textbf{NEOS} & Our Fit \(\chi^2\) & Collab. \(\chi^2\) & \(\Delta m^2\) &
			\(\sin^22\theta\) \\ \hline
			Minimum 1 & 59.73 & \(\sim57.5\) & 1.30 \si{eV^2} & 0.04 \\
			Minimum 2 & 59.60 & 57.5					& 1.73 \si{eV^2} & 0.05 \\
			\hline
		\end{tabular}
		\caption{Comparison of our $\chi^2$ to the values obtained by the collaboration at their quasi-degenerate global minima.}
		\label{tab:NEOS-islands}
	\end{table}

	Due to our incomplete knowledge of the detector response, our fit does not perfectly
	reproduce the results from the collaboration: in particular, we get more local minima in
	our fit, and the location of our global minimum differs from the one obtained by the
	collaboration in \cref{tab:best-fit}. We assume this to be due to the unknown detector
	response, combined with the impact of systematic uncertainties ({\em e.g.,\ }energy scale)
	which are technically difficult to implement outside the collaboration. However, we have
	checked that the differences in $\chi^2$ are relatively small. For example, the
	collaboration obtains two degenerate minima at \((\Delta m^2,\sin^22\theta)=(0.05,
	\SI{1.73}{eV^2})\) and \((0.04,\SI{1.30}{eV^2})\). For these two points of parameter
	space, our fit yields a higher $\chi^2$ by approximately two units, see
	\cref{tab:NEOS-islands}.%

	\subsection {PROSPECT} \label{app:PROSPECT}
	We approximate the detector response from the extensive public data release accompanying
	ref.~\cite{Andriamirado:2020erz}, which contains efficiencies ($\epsilon_s$) and response
	matrices ($R_{ij}^s$) for each segment \(s\) of the detector. For our analysis, we group the
	segments into baseline bins, which we index with \(l\), according to the segment map given
	by the collaboration. The fit is performed using the {\tt GLoBES} \cite{Huber:2004ka,
	Huber:2007ji} libraries. For each bin $l$, the detector mass is set to the sum of the
	relative efficiencies of the segments contained in that bin. Moreover, its effective
	baseline is defined by adding the contribution of each segment, weighted by its relative
	efficiency $w_s=\frac{\epsilon_s}{\sum_s\epsilon_s}$ as
	\begin{align}
		\Braket{\frac{1}{L^2}}_l = \sum_s w_s\Braket{\frac{1}{L^2}}_s \, , && \text{with} &&
		\Braket{\frac{1}{L^2}}_s =\int\d L\,\frac{\rho_s(L)}{L^2} \, .
	\end{align}
	Here, the baseline distribution \(\rho_s(L)\) for each segment \(s\) is constructed by
	randomly selecting points in the reactor core and detector segment to form a
	density histogram (the segment map provided by the collaboration determines the
	detector geometry, while the reactor core is modeled as a cylinder of height
	\SI{51}{cm} and radius \SI{22}{cm}).

	The four-neutrino oscillation probability for a given bin \(l\) and neutrino energy $E_\nu$
	is computed as
	\begin{equation}
		P_{ee}^l\left(\sin^2\theta, \frac{\Delta m^2}{4 E_\nu}\right) =
		1 - \sin^22\theta\cdot\Pi_l\left(\frac{\Delta m^2}{4 E_\nu}\right),
	\end{equation}
	where for each baseline bin $l$ we define
	\begin{equation}
			\Pi_l(q)=
			\left[\sum_s w_s\Braket{\frac{\sin^2(qL)}{L^2}}_s\right] \Bigg/ \Braket{\frac{1}{L^2}}_l
	\end{equation}
	with
	\begin{align}
		\Braket{\frac{\sin^2(qL)}{L^2}}_s&=\int\d L\,\frac{\rho_s(L) \sin^2(qL)}{L^2}.
	\end{align}

	At a given baseline bin $l$ and for an energy bin $i$, the predicted event rates for a set
	of oscillation parameters are proportional to
	\begin{equation}
		N_{(il)} (\sin^22\theta, \Delta m^2) = \sum_j R_{ij}^l \int_{E_j}^{E_{j+1}}\d E_\nu \,
		\left[\sigma_\text{IBD} \frac{\d\Phi}{\d E_\nu}\right](E_\nu) \times P_{ee}^l\left(\sin^22\theta, \frac{\Delta m^2}{4 E_\nu}\right)
	\end{equation}
	where $\dderiv\Phi{E_\nu}$ is the predicted neutrino flux and $\sigma$ is the cross section, while $R^l_{ij}$ stands for the effective energy response
	matrix for bin $l$, given by
	\begin{align}
		R^l_{ij} = \sum_s W_s R_{ij}^s, && \text{where} &&
		W_s=\frac{w_s\Braket{L^{-2}}_s}{\Braket{L^{-2}}_l} \, .
	\end{align}
	We have grouped the energy index $i$ and the baseline index $l$ into a combined index $(il)$
	as a bookkeeping device. The weights $W_s$ account for the relative expected event rates in
	the segments. In principle, these vary as a function of the sterile-neutrino parameters, but
	we assume that these differences are negligible.

	The result from \cref{eq:PROSPECT-pred} is then used to rescale the predicted signal event
	rates by the collaboration in the absence of oscillations ($S^0_{(il)}$, provided in the
	data release), as
	\begin{equation} \label{eq:PROSPECT-pred}
		S_{(il)} = S^0_{(il)}
		\frac{N_{(il)} (\sin^22\theta, \Delta m^2)}{N_{(il)} (\sin^22\theta=0)}.
	\end{equation}
	Once the event rates have been computed, the $\chi^2$ is calculated as in
	ref.~\cite[eq.\ 11]{Andriamirado:2020erz}:
	\begin{equation}
		\chi^2_\text{PROSPECT} = \sum_{l,m}\sum_{i,j}\left(O_{(il)}-S_{(il)}\frac{O_i}{S_i}\right)
		V_{(il)(jm)}^{-1} \left(O_{(jm)} - S_{(jm)}\frac{O_j}{S_j}\right),
	\end{equation}
	where $V$ is the covariance matrix provided by the collaboration,\footnote{Our notation
	reflects the intuition that $V$ is naturally expressed a $160\times160$ matrix, rather than
	as a $10 \times 16 \times 10 \times 16$ tensor, by flattening the rates into a 160-element vector. In particular, the former lends itself to
	inversion more readily than the latter.} $O_{(il)}$ is the observed number of events in
	energy bin $i$ for baseline bin $l$, and we have defined
	\begin{align}
		O_i = \sum_l O_{(il)} && \text{and} && S_i = \sum_l S_{(il)} \, .
	\end{align}

	In \cref{fig:PROSPECT-STEREO-FC-compare} (left), we show our Feldman-Cousins allowed regions
	with statistical fluctuations generated as explained in \cref{app:FC-analysis}; we find
	very good agreement with the results reported by the collaboration.%

	\subsection {STEREO} \label{app:STEREO} %

	For each of the six cells in the STEREO detector, we calculate the effective baseline and
	oscillation probability in a similar fashion to PROSPECT (described in \cref{app:PROSPECT}),
	according to the geometry described in ref.~\cite{AlmazanMolina:2019qul}. We index these
	cells with $l = 1,\ldots,6$. In principle, light produced in a given cell can leak into its
	neighbors, potentially converting a single event of a given energy into multiple events with
	lesser energy in multiple segments. However, we assume that this cross-talk is negligible in
	our analysis.

	The STEREO data set is divided into two phases; we index these phases with $\lambda=\rm I,\,
	{\rm II}$. Here we assume that all cells have the same energy resolution and that the cell
	masses are identical for both phases. The energy-resolution function can be read from
	ref.~\cite[fig.\ 2.19]{Bernard:2019byr} for phase I, and from ref.~\cite[fig.\
	11]{AlmazanMolina:2019qul} for phase II. It is parametrized as
	\begin{equation}
		\begin{aligned}
			\frac{\sigma_\lambda(E)}{E} = a_\lambda + \frac{b_\lambda}{E/\si{MeV}}, &&
			\text{where} &&
			\begin{aligned}
				a_I & = 0.031 , \qquad b_I = 0.059 ; \\
				a_{II} & = 0.043 , \qquad b_{II} = 0.050.
			\end{aligned}
		\end{aligned}
	\end{equation}
	The detector non-linearities can also be found in the same references. The extracted data
	for the non-linear detector responses are fit to quadratic functions,
	\begin{equation}
		f_l^\lambda(E) = c_l^\lambda + d_l^\lambda \frac{E}{\text{MeV}} +
		e_l^\lambda \left(\frac{E}{\text{MeV}}\right)^2 \, ,
		\label{eq:STEREO-fl}
	\end{equation}
	and are applied to the reconstructed energies as \(E_\text{rec}\mapsto E_\text{rec}[1+
	f_l^\lambda(E_\text{rec})]\). The values obtained for the coefficients for each cell are
	provided in \cref{tab:STEREO_nonlinearity} for convenience.
	\begin{table}[t]
		\centering
		\begin{tabular}{|C{0.5cm}|ccc|ccc|}
			\hline & \multicolumn{3}{|c|}{Phase I} & \multicolumn{3}{|c|}{Phase II} \\
			\hline
			$l$ & $c^I_l$ [\%] & $d^I_l$ [\%] & $e^I_l$ [\%] & $c^{II}_l$ [\%] & $d^{II}_l$ [\%] & $e^{II}_l$ [\%] \\
			\hline
			1 & 2.2 & $-0.97$ & 0.092 & $- 0.23$ & 0.25 & $-0.029$	\\
			2 & 1.7 & $- 0.98$ & 0.10 & $- 0.054$ & 0.043 & $-0.010$	\\
			3 & 1.1 & $-0.69$ & 0.074 & $-0.11$ & 0.22 & $-0.027$	\\
			4 & 1.65 & $-0.96$ & 0.01 & 0.15 & $-0.076$ & 0.0026 \\
			5 & 2.0 & $-0.99$ & 0.10 & $-0.10$ & 0.095 & $-0.020$ \\
			6 & 1.2 & $-0.70$ & 0.070 & 0.35 & $-0.38$ & 0.022 \\
			\hline
		\end{tabular}
		\caption{STEREO non-linearity coefficients (\cref{eq:STEREO-fl}) for each cell $l$, as
		obtained from a quadratic fit to the points shown in ref.~\cite[fig.\
		2.19]{Bernard:2019byr} for phase I, and from ref.~\cite[fig.\ 11]{AlmazanMolina:2019qul}
		for phase II. }
		\label{tab:STEREO_nonlinearity}
	\end{table}%

	The energy resolution and non-linear response functions are then combined to build a
	Gaussian response matrix \(R_l^\lambda(E_\text{rec}^i,E_\nu^j)\), used to map events in
	\emph{true} antineutrino energy $E_\nu$ (distributed across 70 bins between 1.85 and 7.95
	MeV) onto the \emph{reconstructed} energy
	$E_\text{rec}$ used in the analysis, listed in
	\cref{sec:exp-reactor}. The corresponding event rates are
	\begin{equation}
	 \begin{split}
		N_{l,i}^\lambda(\sin^22\theta,\Delta m^2) = & \sum_j R_l^\lambda(E_\text{rec}^i,E_\nu^j)
		\times \\
		& \int_{E_\nu^j}^{E_\nu^{j+1}}\d E_\nu \,
		\left[\sigma_\text{IBD} \frac{\d\Phi}{\d E_\nu}\right](E_\nu) \times
		\,P_{ee}^l\left(E_\nu; \sin^22\theta,\Delta m^2\right).
		\end{split}
	\end{equation}

	As mentioned in the main text, the STEREO collaboration has provided their data normalized
	with respect to their best-fit no-oscillation hypothesis for each phase. As such, in order
	to compare our predictions with their data, we must normalize our predictions in the same
	way. To wit, for each phase $\lambda$, the prediction we employ for energy bin $i$ and cell
	$l$ can be written as~\cite{AlmazanMolina:2019qul}:
	\begin{equation}
	\label{eq:pred-defn-stereo}
		S^\lambda_{l,i} = \frac{\left(1+\zeta^\lambda_l\right) \times
		N^\lambda_{l,i}\left(\eta^\lambda_l + \xi^\lambda\right)}
		{\left(1+\bar\zeta^\lambda_l\right) \times N^{0,\lambda}_{l,i}(\bar\eta^\lambda_l+
		\bar\xi^\lambda_l)},
	\end{equation}
	where \(\eta^\lambda_l\) is a pull introduced for the cell-uncorrelated energy-scale
	uncertainty, \(\xi^\lambda\) is its cell-correlated counterpart, and $\zeta_l^\lambda$ is a
	pull for the signal-normalization uncertainty. Here, \(N^{0,\lambda}\) is the prediction for
	the no-oscillation hypothesis for the fixed (optimized) pulls \(\bar\eta^\lambda_l,\bar\xi^
	\lambda_l,\bar\zeta^\lambda_l\) provided in ref.~\cite[fig.\ 31]{AlmazanMolina:2019qul}.

	The $\chi^2$ function for STEREO is computed by adding the contributions of data from the
	two phases. Following ref.~\cite[eqs.\ (15, 19)]{AlmazanMolina:2019qul}, we include a set
	of free flux normalizations for each bin $i$ ($\phi_i$, common to the two
	phases)\footnote{Note that while the ten lowest-energy bins are identical for the two
	phases, there is an additional high-energy bin for the phase-II data, ({\em i.e.},
	$\NE^{II}=\NE^I +1$). As such, the phase II analysis depends on an additional flux nuisance
	parameter $\phi$ to which phase I is insensitive.} as well as a relative normalization
	uncertainty between the two phases ($\Phi^{\rm I}$, common to all bins). The final $\chi^2$
	function is obtained after minimization over all nuisance parameters:
	\begin{equation}
	\label{eq:chi2-stereo}
		\begin{aligned}
			\chi^2_\text{STEREO} =
			\min_{\substack{\eta^{I,II}_l;\;\xi^{I,II}\\ \zeta_l^{I,II};\;\phi_i;\;\Phi}}
			\Bigg\{ &\sum_{l=1}^{\Nc}\sum_{i=1}^{\NE^I}
			\left(
			\frac{O^I_{l,i}-\Phi^I\phi_i S^I_{l,i}(\eta^I_l+\xi^I, \zeta^I_l)}{\sigma^I_{l,i}}
			\right)^2 \Bigg. \\
			& + \sum_{l=1}^{\Nc} \sum_{i=1}^{\NE^{II}} \left(
			\frac{O^{II}_{l,i}-\phi_i S^{II}_{l,i}(\eta^{II}_l+\xi^{II},\zeta^{II}_l)}
			{\sigma^{II}_{l,i}}
			\right)^2 \\
			&+ \Bigg.\sum_{\substack{l=1\\ \lambda=I,II}}^{\Nc} \left(
			\frac{\eta_l^\lambda}{\sigma_\eta^\lambda}\right)^2 +
			\left(\frac{\zeta_l^\lambda}{\sigma_\zeta^\lambda}\right)^2 +
			\sum_{\lambda=I,II}\left(\frac{\xi^\lambda}{\sigma_\xi^\lambda}\right)^2
			\Bigg\}.
		\end{aligned}
	\end{equation}
	Here, \(O_{l,i}^\lambda\) are the observed data points in baseline bin $l$ and energy bin
	$i$ \cite{Almazan:2018wln,AlmazanMolina:2019qul} while $\sigma_{l,i}^\lambda$ are the
	statistical errors on the data. The uncertainties of the nuisance parameters are:
	\begin{equation}
		\begin{aligned}
			\sigma^I_\eta	=1.06\%, && \sigma^I_\xi = 0.35\%, && \sigma^I_\zeta=1.18\%, \\
			\sigma^{II}_\eta=1.02\%, && \sigma^{II}_\xi=0.30\%, && \sigma^{II}_\zeta=1.18\% ,
		\end{aligned}
	\end{equation}
	while both $\phi_i$ and $\Phi^I$ are left completely \emph{unconstrained} in the fit.

	Finally, we make the transformations
	\begin{align}
		\alpha_l^{\lambda} = \eta^{\lambda}_l+\xi^{\lambda}, && \Phi^I(1+\zeta^I_l) = 1+\beta^I_l,
	\end{align}
	and for consistency \(\beta^{II}_l=\zeta^{II}_l\). This isolates the dependence of
	\(\xi^\lambda \), \(\Phi^I\) in the quadratic pull terms, which we can analytically
	minimize. We numerically minimize the remaining pulls with a non-linear conjugate-gradient
	algorithm \cite{Fletcher:2000}.%

	\begin{figure}
		\centering
		\includegraphics[width=0.45\textwidth]{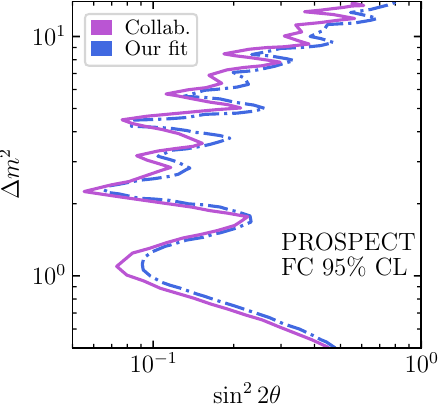}
		\hfill
		\includegraphics[width=0.45\textwidth]{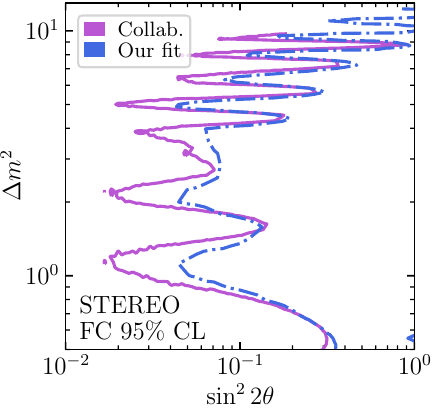}
		\caption{Comparison of our 95\% FC curve for PROSPECT (left) and STEREO (right), both
		evaluated with \(10^4\) pseudo experiments, to the result of the collaboration
		\cite{Andriamirado:2020erz,STEREO-HEP-data}.
		}
		\label{fig:PROSPECT-STEREO-FC-compare}
	\end{figure}

	Using our $\chi^2$ implementation, we are able to reproduce the results of the collaboration
	to a reasonable degree of accuracy, both for the $\Delta \chi^2$ as well as for the 95\%
	\({\rm CL_s}\)\footnote{See \cite{Qian:2014nha} for an explanation of the \({\rm CL_s}\)
	procedure.}, curves published in their data release \cite{STEREO-HEP-data}. A comparison of
	our FC 95\% C.L. curve is shown in \cref{fig:PROSPECT-STEREO-FC-compare} (right). While the
	shape of our exclusion curve is similar to the result of the collaboration, we do find a
	general reduction in our calculated exclusion. We think, however, that the level of
	agreement achieved is reasonable, given that:
	\begin{enumerate}
		\item The $\chi^2$ is quite flat in the low-$\Delta m^2$ region. Consequently, small
			numerical differences --- in either the low-$\Delta m^2$ region or around the global
			minimum --- can lead to sizeable shifts in the locations of contours of constant $\Delta
			\chi^2$.
		\item Our minimization procedure has finite resolution. If the global minimum is missed
			during the minimization procedure, then this would manifest as an apparent reduction in
			sensitivity over the entire parameter space. Increasing the resolution beyond the
			current level would make our simulation computationally too expensive.
		\item Our detector response is only approximate. We use cell- and phase-specific
			descriptions of the non-linearity, but assume a common resolution for all cells for a
			given phase. This may affect the high-$\Delta m^2$ region in particular.
	\end{enumerate}

	Finally note that, when generating pseudo data for STEREO, for some of the bins in the
	high-energy tail, we sometimes obtain negative fluctuations. Whenever that happens, we
	simply throw these away and redraw fluctuations for all bins.

	\subsection {Neutrino-4} \label{app:nu4}
	Our analysis of Neutrino-4 is based on ref.\cite{Coloma:2020ajw}. As outlined in \cref{sec:exp-reactor}, the event rates are binned in $L/E$ and grouped
	into 27 bins. In our analysis, we use the first 19 bins of such a data set, shown as blue
	triangles in fig.~47 in v2 of the preprint of ref.~\cite{Serebrov:2020kmd}.

	The collaboration performed their analysis in terms of the ratio of events in each bin to
	the total rates in all bins, as in \cref{eq:R-nu4}. The predicted event rates in each bin
	(in the absence of statistical fluctuations) can be computed as\footnote{Note the slightly
	different definition in the denominator with respect to ref.~\cite{Coloma:2020ajw}.
	While the expression used in ref.~\cite{Coloma:2020ajw} is theoretically correct, in
	practice the average over different bins is not perfect and can lead to small differences
	in the region of small $\Delta m^2$. Thus, here we adopt the exact expression instead.}
	\begin{equation} \label{eq:nu4-pred}
		R_i^{\rm pred} =
		\frac{1 - \sin^22\theta \Braket{ \sin^2\frac{\Delta m^2 \, L}{4E} }_i }
		{1 - \sin^22\theta \frac1N\sum_{i=1}^N\Braket{\sin^2\frac{\Delta m^2\,L}{4E}}_i } \, ,
	\end{equation}
	where the average over each bin $i$ is indicated by $\braket{\,\cdot\,}_i$ and takes
	into account both energy and baseline uncertainties.

	The fit to Neutrino-4 data is performed with a simple Gaussian $\chi^2$ definition:
	\begin{equation}
		\chi^2_{\rm Neutrino-4} =
		\sum_{i=1}^{19} \frac{(R_i^{\rm obs} - R_i^{\rm pred})^2}{\sigma_i^2} \, ,
	\end{equation}
	where $\sigma_i$ corresponds to the statistical uncertainty read off fig.~47 of
	ref.~\cite{Serebrov:2020kmd}, while $R_i^{\rm obs}$ and $R_i^{\rm pred}$ refer to the
	observed and predicted ratios in each bin.

	Once the theoretical prediction has been computed for each bin, we generate statistical
	fluctuations in each bin according to a Gaussian centered at the theoretical prediction,
	$\mathcal{N}(R_i^{\rm pred}, \sigma_i)$, where $\sigma_i$ is the uncertainty in each bin
	extracted from the same figure\footnote{In order to reproduce the results in
	ref.~\cite{Serebrov:2020kmd} better, we scale down their uncertainties by 3\%.} in
	ref.~\cite{Serebrov:2020kmd}. %
	
	Let us mention that Neutrino-4's treatment of the detector energy resolution
	has been criticized in refs.~\cite{Danilov:2018dme,Danilov:2020rax,Giunti:2021iti}.
	In particular, a proper treatment of the energy resolution may reduce the reported
	signal significance by approximately 0.5$\sigma$~\cite{Giunti:2021iti}, and require
	even larger values for the mixing angle. Additionally, the authors of
	ref.~\cite{PROSPECT:2020raz} have claimed that unaccounted correlations between
	energy and efficiency may induce an oscillation-like signature at Neutrino-4. In our
	analysis we do take into account finite energy (as well as spatial) resolution, see
	\cref{eq:nu4-pred}. In choosing the value for the resolution we follow our general
	strategy, to reproduce as close as possible the official oscillation results
	published by the collaboration. Specifically, the average in each bin is performed
	using an effective energy uncertainty $\Delta E_\mathrm{eff} = E \sqrt{(\Delta L/L)^2
	+ (\Delta E/E)^2}$, where $\Delta E = 500~\mathrm{keV}$ and $\Delta L =
	0.48~\mathrm{m}$.

	\subsection {Solar Neutrinos} \label{app:solar}%

	The solar-neutrino analysis performed here is based on the simplified $\chi^2$
	construction from ref.~\cite{Goldhagen:2021kxe}, which in turn is based on the global fit
	from ref.~\cite{Esteban:2020cvm}. While we refer the interested reader to the
	aforementioned references for details, here we provide a brief explanation of the solar
	analysis for completeness.

	Our fit considers four data points as in ref.~\cite{Goldhagen:2021kxe}, corresponding to
	the four oscillation probabilities:
	\begin{equation} \label{eq:solar-obs}
		r = (P_{ee}^{\rm LE}, \, P_{ee}^{\rm HE} ,\,
		P_{ex}^{\rm LE}, \, P_{ex}^{\rm HE}) .
	\end{equation}
	Here, $P_{ee}$ is the electron-neutrino survival probability and $P_{ex}=P_{e\mu}+
	P_{e\tau}$ is the transition probability of electron neutrinos to the other active
	neutrino flavours. The indices ${\rm LE}$ (low energy) and ${\rm HE}$ (high energy) refer
	to the energy regions below and above the MSW resonance, respectively.

	Under the same set of assumptions adopted in \cite{Goldhagen:2021kxe}, the probabilities
	in \cref{eq:solar-obs} only depend on the angles \(\theta_{12}\), \(\theta_{13}\),
	\(\theta_{14}\). Since it has been shown in ref.~\cite{Kopp:2013vaa} that the
	determination of $\theta_{13}$ is unaffected by the presence of a sterile neutrino,
	throughout this work we will fix the value of $\theta_{13}$ to the three-neutrino best-fit
	point from ref.~\cite{Esteban:2020cvm}, $\sin^2\theta_{13} = 0.0223$. Moreover, in
	ref.~\cite{Goldhagen:2021kxe} it was also shown that the most conservative bound for
	$\theta_{14}$ is achieved for $\sin^2\theta_{12} = 0.3125$. Thus, this will also be fixed
	throughout, for simplicity (we have explicitly checked that allowing
	$\theta_{12}$ to vary has a negligible impact on our results).

	The $\chi^2$ function for solar data is computed as:
	\begin{equation}
		\label{eq:chi2-solar}
		\chi^2_{\rm Solar} = \sum_{r,s} (O_r - P_r) V^{-1}_{rs}(O_s-P_s) \,.
	\end{equation}
	where $O_r$ and $P_r$ stand for the observed and predicted values for the probabilities in
	\cref{eq:solar-obs}, respectively, while $V^{-1}_{rs}$ denotes the inverse covariance
	matrix, see \cref{eq:cov-solar} below. The observations $O_r$, as well as the
	corresponding uncertainties $\sigma_r$ and correlation matrix $\rho$, are taken from
	ref.~\cite[tab.~1]{Goldhagen:2021kxe}. In this work we use the values for the GS98 solar
	model \cite{Vinyoles:2016djt}, corresponding to the upper half of the table. Let us define
	the relative covariance	matrix for the observations as
	\begin{equation}
		S_{rs} = \rho_{rs} \frac{\sigma_r}{O_r} \frac{\sigma_s}{O_s}
	\end{equation}
	(no sum over repeated indices). Following ref.~\cite{Goldhagen:2021kxe}, a good
	approximation to the full solar-neutrino fit is obtained by splitting the inverse covariance
	matrix $V^{-1}_{rs}$ into a theoretical and an experimental part, as
	\begin{align} \label{eq:cov-solar}
		V^{-1}_{rs} = \frac{S^{-1}_{rs}}{2}
		\left[ \frac{1}{O_r} \frac{1}{O_s} + \frac{1}{P_r} \frac{1}{P_s} \right]
		= \frac12 \left[ X^{-1}_{rs} + S^{-1}_{rs}\frac{1}{P_r}\frac{1}{P_s} \right] \, ,
		&& \text{where} && X_{rs} = \rho_{rs} \sigma_r \sigma_s
	\end{align}
	(no sum over repeated indices), and $S^{-1}$ is the inverse of $S$. Using the inverse
	covariance matrix defined in \cref{eq:cov-solar} the $\chi^2$ for solar data can be computed
	as in \cref{eq:chi2-solar}.

	Regarding generation of pseudo data, note when $P_r = O_r$, we have $V_{rs} = X_{rs}$ and
	the dependence on $P_r$ drops out (see \cite[eq.\ (2.12)]{Goldhagen:2021kxe}). Therefore,
	we use the Cholesky decomposition of $X$ in order to generate the statistical fluctuations
	for the pseudo data, following the procedure outlined in \cref{app:MC}.%

	\section {Details of the Feldman-Cousins Analysis} \label{app:FC-analysis}
	\subsection {Treatment of nuisance parameters}
	\label{app:sys}%

	When generating pseudo data, in addition to statistical fluctuations on the event rates, we
	need to include fluctuations on the nuisance parameters as well. This means that the pseudo
	data should be generated using a set of ``true'' nuisance parameters, $\xi_\text{true}$, as
	the mean for \(\xi\). The question is what value should be taken for $\xi_\text{true}$.
	Within a frequentist approach, {\em a priori} one should scan all possible values of $\xi$
	and study the distribution of the test statistic accordingly (that is, treat the nuisance
	parameters as if they were additional parameters of the model). However, in practice this
	is often not possible and a choice has to be made regarding the assumed values for
	$\xi_\text{true}$.

	For all reactor experiments considered in this work, we will leave the overall normalization
	of the event rates completely free; as such, we will \emph{not} consider fluctuations of
	these free pulls, fixing them to the values that provide the best fit to the observed data.
	We do so, because a free normalization implies that no prior information on that
	parameter is being used. Therefore, since the only data to consider are those measured
	by the experiment, it seems plausible to use them in order to determine the flux
	normalization. In other words: we allow each experiment to ``choose'' their preferred value
	of the overall normalization for the event rates, given that we are only interested in the
	observation of an oscillatory signal in the spectrum. Note that this is done independently
	for each set of assumed ``true'' oscillation parameters, $\sin^22\theta, \Delta m^2$.

	For the rest of the nuisance parameters considered here, prior measurements are typically
	available and can be used to determine the relevant range of values to consider when
	generating the fluctuations. Thus, for each nuisance parameter we will generate Gaussian
	fluctuations around the best-fit values from previous measurements ({\em i.e.,\ }the
	mean value as reported by the collaboration), taking the reported uncertainties as the
	width of the Gaussian.%

	\subsection {Generation of pseudo data}
	\label{app:MC}%

	We use the following procedure to generate pseudo data in our analysis:
	\begin{enumerate}[(i)]
		\item choose the parameters \(\Delta m^2\), \(\sin^22\theta\) for the hypothesis of new
			oscillations;
		\item calculate the normalization-related nuisance parameters $\{\eta^\text{min}\}$ that
			provide the best fit to the observed data for the assumed hypothesis;
		\item generate fluctuations for the nuisance parameters subject to prior constraints,
			taking as central values their reported best-fit values $\overline\xi_k$, {\em i.e.},
			\begin{equation}\label{eq:gen-pulls}
			\xi_k^\text{pseudo} = \overline\xi_{k} + \sigma_{\xi_k} \delta_k \,,
			\end{equation}
			 where $\delta_k$ are standard-normal fluctuations and $\overline\xi_k$ are the values to which the pulls are constrained by the penalty terms;
		\item compute the prediction for the assumed hypothesis, using the fluctuated nuisance
			parameters and calculated flux normalizations, and set this as the mean for the
			generation of pseudo data, {\em i.e.,} for each bin $i$ set
			\begin{equation}\label{eq:Dmean}
				\overline{D}_i^\text{pseudo}=
				P_i\left(\Delta m^2,\sin^22\theta;\{\xi^\text{pseudo}\},\{\eta^\text{min}\}\right);
			\end{equation}
		\item if the \(\chi^2\) function contains a covariance matrix \(V\), first recalculate
			(if necessary) by replacing all data with \(\overline{D}_i^\text{pseudo}\) in \(V\);
		\item calculate the Cholesky decomposition \(L\) of \(V^\text{pseudo}\) (note if
			\(\chi^2\) does not contain a covariance matrix, then \(L\) is diagonal with the
			standard deviations of each bin as the diagonal entries);
		\item inject fluctuations around the pseudo-data mean with \(L\), {\em i.e.,}
			\begin{equation}\label{eq:Dpseudo}
				D_i^\text{pseudo} = \overline{D}_i^\text{pseudo} + \sum_j L_{ij}\delta_j,
			\end{equation}
			where \(\delta_j\) are standard-normal fluctuations.
	\end{enumerate}%

	\subsection{Comment on the \texorpdfstring{$\chi^2$}{χ²} distribution related to randomized
	nuisance parameters}

	Here we provide some justification for the way we generate pseudo data in the presence of
	constrained nuisance parameters. For this aim we consider the simplified case of a single
	pull $\xi$ describing a normalization uncertainty. Without loss of generality we set
	$\overline\xi = 0$, such that \cref{eq:gen-pulls,eq:Dmean} become:
	\begin{align}
		\xi^\text{pseudo} &= \sigma_{\xi} \, \delta \,, \label{eq:xi1}\\
		\overline{D}_i^\text{pseudo} &= (1+ \xi^\text{pseudo}) P_i(\theta)  \,, \label{eq:xi2}
	\end{align}
	where $\theta$ denotes generic model parameters.  For simplicity, let us also assume that
	the data is a priori uncorrelated with variances $\sigma_i^2$ and \cref{eq:Dpseudo} becomes
	\begin{equation}
		D_i^\text{pseudo} = \overline{D}_i^\text{pseudo} + \sigma_i \delta_i \,. \label{eq:xi3}
	\end{equation}
	Then we obtain for our test statistic for the pseudo data:
	\begin{align}\label{eq:chi2pull}
		\chi^2_{\rm pseudo} = \min_\xi\left[ \sum_{i=1}^N \frac{[D_i^{\rm pseudo} - (1+\xi)P_i(\theta)]^2}{\sigma_i^2} +
		\frac{\xi^2}{\sigma_\xi^2} \right] \,.
	\end{align}

	Let us now prove that for fixed parameters $\theta$, the test statistic $\chi^2_{\rm
	pseudo}$ is distributed as a $\chi^2$ with $N$~dof. First, it follows from
	\cref{eq:xi1,eq:xi2,eq:xi3} that $D_i^{\rm pseudo}$ are multivariant Gaussian variables with
	mean and covariances of
	\begin{equation} \label{eq:meanS}
		\langle D_i^{\rm pseudo} \rangle = P_i(\theta)\,,\quad\text{Var}(D_i^{\rm pseudo},D_j^{\rm
		pseudo}) \equiv S_{ij} = \delta_{ij} \sigma_i^2 + P_i(\theta)P_j(\theta) \sigma_\xi^2 \,,
	\end{equation}
	with Var$(X_i,X_j) = \langle X_i X_j\rangle - \langle X_i\rangle \langle X_j\rangle$ and
	$\delta_{ij}$ is the Kronecker delta. On the other hand, \cref{eq:chi2pull} has the standard
	form of a pull-$\chi^2$, which is known\footnote{See ref.~\cite{Fogli:2002pt} for an
	explicit proof.} to be equivalent to $\chi^2_{\rm pseudo} = \sum_{ij}[D_i^{\rm pseudo} -
	P_i(\theta)]S_{ij}^{-1}[D_j^{\rm pseudo} - P_j(\theta)]$ with $S_{ij}$ given in
	\cref{eq:meanS}. Hence, by a suitable variable transformation (involving the diagonalization
	of $S$) one can write $\chi^2_{\rm pseudo} = \sum_{i=1}^N \zeta^2_i$, with $\zeta_i$ being
	$N$ independent standard-normal random variables, and therefore $\chi^2_{\rm pseudo}$
	follows a $\chi^2$ distribution with $N$ dof.

\emph{Remark 1}: This proof generalizes in a straight-forward way if there is a nontrivial correlation matrix between the data points, and to the case of several pulls, as long as they enter the prediction linearly.

\emph{Remark 2:} In this proof it has been essential to use the reported best-fit value
	$\overline{\xi}$ as central value to fluctuate pull, as in \cref{eq:gen-pulls}. (Remember
	that we have set $\overline\xi = 0$ in this section, without loss of generality.) If for
	generating the random pulls a mean value different from $\overline{\xi}$ was used, the
	pseudo-data would be biased and $\chi^2_{\rm pseudo}$ would follow a non-central $\chi^2$
	distribution.

	\subsection {MC uncertainty of FC confidence intervals} \label{app:FC-error}%

	We describe how we estimate the uncertainty on our FC confidence regions due to the
	finite size of the MC sample. Consider a point in the parameter space, which has a
	true $p$-value $p$. Then for a total of $N$ MC trials, the number $n$ of pseudo-data
	sets with $\Delta\chi^2_\text{pseudo} > \Delta\chi^2_\text{obs}$ is distributed as a
	binomial distribution, with parameters $N$ and $p$. Hence, we can define a \(p\)-value
	distribution via $f(q) = N B\left(n;N,p\right)$, where $q=n/N$ and \(B(\,\cdot\,;\,\cdot
	\,,\,\cdot\,)\) is the binomial PDF. An \(\alpha=99\%\) confidence interval \([q_0,q_1]\)
	for this $p$-value can be defined as
	\begin{align}\label{eq:uncert-boundaries}
		F(q_0)=\frac{1 - \alpha}{2} \,,\qquad
		F(q_1) = \frac{1 + \alpha}{2} \,,
	\end{align}
	where \(F(\,\cdot\,)\) is the CDF of \(f(\,\cdot\,)\). For small \(p\)-values, the
	distribution is skewed. Note that since the distribution $f$ is discrete, the relations in
	\cref{eq:uncert-boundaries} are only approximate; they become exact in the large-$N$ limit.

	The MC approximation for the \(p\)-value is \(n/N\). We take this as the \(p\)
	parameter for the binomial distribution ({\em i.e.,\ }we assume the MC approximation is
	exact). To obtain the 99\%-confidence uncertainty bands for a confidence region, we
	create a $p$-value map in the $\sin^22\theta$--$\Delta m^2$ plane. Then we interpret
	the area between the contours at \(q_0\) and \(q_1\) as the 99\% confidence
	uncertainty on the $\beta$~CL contour with $\beta=1-p$.

\begin{figure}
		\centering
		\includegraphics[width=\textwidth]{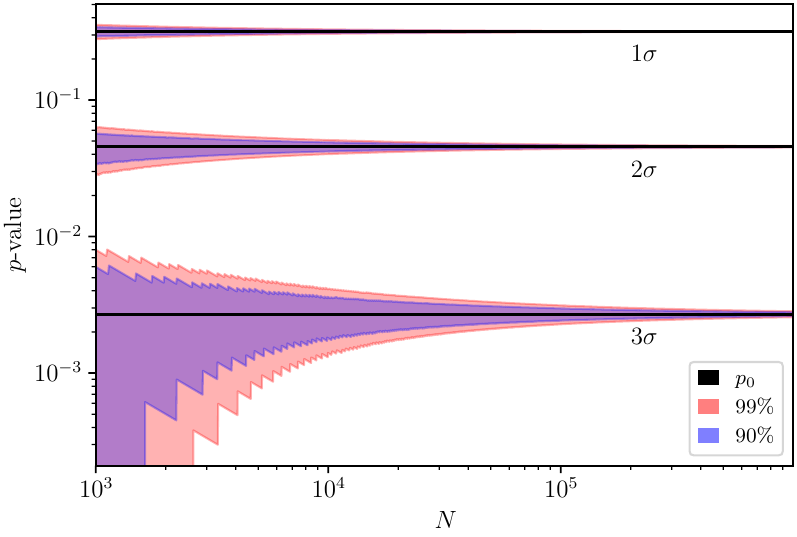}
		\caption{90\% and 99\% uncertainty on the $p$-value for $p = 0.32$ ($1\sigma$), 0.045 ($2\sigma$), and 0.0027 ($3\sigma$) as a function of the MC sample size $N$.
		}
		\label{fig:FC-error}
	\end{figure}%

	Note that this prescription, and in particular \cref{eq:uncert-boundaries}, is independent
	of the specific experiment. It follows from the properties of the binomial distribution and
	depends only on the CL $\beta = 1-p$, the MC sample size $N$, and the uncertainty $\alpha$.
	We plot some numerical values in \cref{fig:FC-error}. The discontinuities visible in the
	bands for $3\sigma$~CL are due to the discrete nature of the binomial distribution. Note
	that for a MC sample size of $10^3$ it is not possible to calculate a meaningful $3\sigma$
	confidence region at the accuracy of $\alpha = 0.99$ or 0.9, as follows from $q_0 = 0$ for
	these parameters.%

\end{subappendices}
\bibliographystyle{JHEP}
\bibliography{Doc_Bib}
\end{document}